2014

# QCD and Hadron Physics

Summary of the DNP Town Meeting
Temple University, 13-15 September 2014

Stanley J. Brodsky, Abhay L. Deshpande, Haiyan Gao, Robert D. McKeown, Curtis A. Meyer, Zein-Eddine Meziani, Richard G. Milner, Jianwei Qiu, David G. Richards, Craig D. Roberts

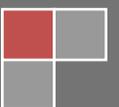

2/16/2015

# Contents

# 1. Executive Summary

## 1.1 Background

This White Paper presents the recommendations and scientific conclusions from the Town Meeting on *QCD and Hadronic Physics* that took place in the period 13-15 September 2014 at Temple University as part of the NSAC 2014 Long Range Planning process. The meeting was held in coordination with the Town Meeting on *Phases of QCD* and included a full day of joint plenary sessions of the two meetings. A total of 244 physicists registered for the joint meetings, with 136 for *QCD and Hadron Physics* and 108 for *Phases of QCD*. The meeting agenda is included in the Appendix.

The goals of the meeting were to report and highlight progress in hadron physics in the seven years since the 2007 Long Range Plan (LRP07),[1] and present a vision for the future by identifying the key questions and plausible paths to solutions which should define our next decade. In defining the priority of outstanding physics opportunities for the future, both prospects for the short ($\sim$ 5 years) and longer term (beyond 10 years) are identified together with the facilities, personnel and other resources needed to maximize the discovery potential and maintain U.S. leadership in hadronic physics worldwide. The Town Meeting program consisted of five major themes: Hadron structure at short distances; Hadron structure at long distances; Hadron spectroscopy; QCD and nuclei; and Theory. The joint sessions with *Phases of QCD* were constituted from overview talks on Theory, QCD and Hadron Physics, Phases of QCD; sessions discussing the discovery potential of an Electron-Ion Collider and the Community's need for this facility; and a session dedicated to considering input from the *Workshop on High Performance Computing* and the Town Meeting on *Education and Innovation*.

In the remainder of this introductory summary, we detail the recommendations and their supporting rationales, as determined at the Town Meeting on *QCD and Hadron Physics*, and the endorsements that were voted upon.

The larger document is organized as follows. Section 2 highlights major progress since the 2007 LRP. It is followed, in Section 3, by a brief overview of the physics program planned for the immediate future. Finally, Section 4 provides an overview of the physics motivations and goals associated with the next QCD frontier: the Electron-Ion-Collider.

We would like to note before continuing that in preparation for this town meeting, and previously in connection with the 2012 *Report to NSAC on Implementing the 2007 Long Range Plan*,[2] numerous excellent whitepapers were prepared by members of our community, such as those concerning JLab12[3] and an EIC,[4] and others available at the Town Meeting website ( [phys.cst.temple.edu/qcd/](phys.cst.temple.edu/qcd/) ). We do not attempt to rewrite or revise those documents. Instead we will draw upon them wherever appropriate.

## 1.2 Recommendations and Rationale

The QCD and Hadron Physics Town Meeting, with 136 registered participants, took place at Temple University of over three days: 13-15 September 2014. The following four recommendations are the outcome of that meeting.

> **RECOMMENDATION I**
>
> **With highest priority, we recommend both completion of construction and full operation of the 12 GeV CEBAF at the Thomas Jefferson National Accelerator Facility, along with targeted instrumentation investments, such as the SoLID and MOLLER projects.**

Understanding the fundamental nature of hadrons and nuclei in terms of QCD, the strong interaction piece of the Standard Model, is a central goal in the field of nuclear physics. The last decade has seen the development of new experimental and theoretical tools to quantitatively study the nature of confinement and the structure of hadrons comprised of light quarks and gluons. Together these will allow both the spectrum and the structure of hadrons to be elucidated in unprecedented detail. The 12-GeV upgrade of the Continuous Electron Beam Accelerator Facility (CEBAF) at Jefferson Lab will provide new capabilities that enable an experimental program with substantial discovery potential which can be employed to address these and other important topics in nuclear, hadronic and electroweak physics. For example, exotic hadrons, which might signal novel excitations of the gluon field, will be sought: their discovery would provide a new doorway to explore hadronic matter. In addition, multidimensional images of the nucleons will be produced, which hold great promise to reveal the dynamics of the key underlying degrees of freedom in QCD. In particular, these multidimensional distributions open a new window on the elusive spin content of the nucleon through observables that are directly related to the orbital angular momenta of quarks and gluons. Moreover, analytical and computational techniques in non-perturbative QCD now promise to provide insightful and quantitative predictions that can be meaningfully confronted with, and elucidated by, forthcoming experimental data. The upgraded facility will also enable experimental studies of fundamental short-distance properties in nuclei, which will provide a quantitative understanding of nuclear properties and their relation to the distribution of quarks and gluons in nuclei. Furthermore, the development of extremely high intensity, highly polarized and extraordinarily stable beams of electrons provides innovative opportunities for probing (and extending) the Standard Model, both through parity violation studies and searches for new particles.

The 12-GeV CEBAF is accompanied by new detector upgrades, such as CLAS12, SHMS, and an entirely new experimental hall featuring searches for QCD exotic states (GlueX), which provide a wide range of novel capabilities. However, while the currently envisioned program includes both high rate capability and large acceptance devices, there is no single device that is capable of handling high luminosity ($10^{36}$-$10^{39}$ $cm^{-2}s^{-1}$) over a large acceptance. Therefore, the capabilities of the 12 GeV upgrade will not have been fully exploited unless a large acceptance high luminosity device is constructed. The SoLID (Solenoidal Large Intensity Detector) program is designed to fulfill this need. SoLID is made possible by developments in both detector technology and simulation accuracy and detail that were not available in the early stages of planning for the 12 GeV program. The spectrometer is designed with a unique capability for reconfiguration in order to optimize capabilities for either Parity-Violating Deep Inelastic Scattering (PVDIS) or Semi-Inclusive Deep Inelastic Scattering (SIDIS), and threshold production of the $J/\Psi$ meson. Recent years have also seen proposal development for the MOLLER experiment, which would perform high precision tests of the Standard Model in parity-violating electron-electron scattering. Such an experiment offers a unique opportunity to test the Standard Model. It complements the capabilities of the upgraded Large Hadron Collider and thereby adds enormously to the physics reach and impact of the 12 GeV CEBAF.

The potential of the CEBAF upgrade led the recent 2013 NSAC subcommittee report, *Implementing the 2007 Long Range Plan*, to conclude that "The 12 GeV CEBAF Upgrade when completed will transform Jefferson Lab into a remarkable facility that will provide a number of outstanding opportunities to understand the nature of QCD, the nucleon, and the nucleus. In addition, the unprecedented combination of high intensity, high energy, high longitudinal polarization and beam stability yields unique capabilities that make possible a new generation of experiments probing the nature of fundamental forces in the very early universe." In order to fully realize the scientific potential of the 12-GeV CEBAF, strong support of researchers at laboratories and universities is essential.

> **RECOMMENDATION II**
>
> **A high luminosity, high-energy polarized Electron Ion Collider (EIC) is the highest priority of the U.S. Nuclear Physics QCD community for new construction after FRIB.**

The Electron Ion Collider (EIC) will image the gluons and sea quarks in the proton and nuclei with unprecedented precision and probe their many-body correlations in detail, providing access to novel emergent phenomena in QCD. It will definitively resolve the proton's internal structure, including its spin, and explore the QCD frontier of ultra-dense gluon fields in nuclei at high energy. These advances are made possible by the EIC's unique capability to collide polarized electrons with polarized protons and light ions at unprecedented luminosity over a broad energy range and electrons with heavy nuclei at high energy.

By precisely imaging gluons and sea quarks inside the proton and nuclei, the EIC will address some of the deepest fundamental and puzzling questions nuclear physicists ask:

- How are the gluons and sea quarks, and their spins, distributed in configuration- and momentum-space inside the nucleon? What is the role of the orbital motion of sea quarks and gluons in building the nucleon spin?
- What happens to the gluon density in nuclei at high energy? Does it saturate? How does this phenomenon manifest itself in nucleons?
- How does the nuclear environment affect the distributions of quarks and gluons and their interactions in nuclei? How does nuclear matter respond to a fast moving color charge passing through it? How do quarks dress themselves to become hadrons?

A full understanding of QCD, in a regime relevant to the structure and properties of hadrons and nuclei, demands a new era at the EIC of precision measurements that can probe the full complexity of these basic, compound objects. Theoretical advances over the past decade have resulted in the development of a powerful formalism that provides quantitative links between such measurements and the above questions that physicists are trying to answer. Another important advance in recent years is the increasing precision and reach of *ab initio* calculations performed with lattice QCD techniques. Using experimental data from an EIC, physicists will be able to undertake the detailed comparative study between experimental measurements and predictions made by continuum- and lattice-QCD theory, as well as elucidate the many aspects of hadron and nuclear structure whose investigation still requires more phenomenological theoretical methods.

Accelerator Technology has recently developed so that an EIC with the versatile range of kinematics, beam species and polarization that are crucial to address the above questions, can now be constructed at an affordable cost. Realizing the EIC will be essential in order to maintain U.S. leadership in the important fields of nuclear and accelerator physics.

**RECOMMENDATION III**

**We recommend strong support for other existing facilities, such as the polarized proton facility at RHIC, university-based laboratories, and the scientists involved in these efforts, in order to guarantee the effective utilization of such resources for continued scientific leadership and discovery, and for educating the next generation of nuclear scientists in the USA.**

The discovery potential of the experimental and theoretical study of strong-interaction phenomena and its educational impact are both greatly enhanced by a diverse range of programs which either capitalize upon collaborations that join researchers at the major facilities in common efforts, or invest in research and collaborations at other user facilities or university-based laboratories.

A primary example is the RHIC-Spin program, which joins scientists from national labs and universities, in the USA and abroad, in an investigation of the proton's spin structure using strongly interacting probes. This approach provides a critical complement to the lepton scattering program, and together they provide important contributions in our quest to discover how the spin and orbital angular momentum of the gluons and quarks within the proton combine to produce the value of ½ that characterizes nature's most important fermion.

Making use of the versatility of the proton-proton facility at RHIC, the collaborations have obtained direct evidence for a gluonic component of the proton spin, and sea quark polarizations through electroweak interactions. With new forward instrumentation and continued support for polarized beam operations at RHIC, planned measurements will improve the precision, extend the kinematic reach of measurements sensitive to gluon polarization, and explore unique transverse spin phenomena.

Other examples on a smaller scale are also readily identified: a polarized Drell-Yan program at FermiLab, which will present exceptional opportunities to measure nucleon valence and sea quark spin distributions with high precision; the High Intensity Gamma-Ray Source (HIGS) at the Triangle Universities Nuclear Laboratory (TUNL), which is the world's most intense polarized γ-ray source with wide applications in low-energy hadron physics; and US leadership of programs at numerous other facilities worldwide.

These efforts, others like them, and the scientists involved add enormously to the diversity of material and intellectual resources that can be focused upon both the problem of unraveling the most important features of the Standard Model and exposing phenomena that lay outside its domain and used in training a new generation of nuclear scientists.

**RECOMMENDATION IV**

**We recommend that support for the hadron theory program be increased, in a balanced manner and in proportion to new and continuing investment in experiment. This will both guarantee that all aims of the existing program can most rapidly be achieved and secure a promising future for the next generation of nuclear scientists and the nation. Given the breadth of the hadron physics enterprise, this program must necessarily be multifaceted and**

**capable of reacting quickly to the new opportunities that innovative experiment and creative theory will reveal.**

Theoretical hadron physics in the USA is an eclectic enterprise, with creative, world-leading efforts in computation, phenomenology and theory. Diversity is a great strength of this enterprise, which provides the capacity to interpret and understand empirical observations, and to guide and stimulate new endeavors, both those aimed at completely exposing the content of the Standard Model and those seeking to discover and understand the phenomena that most certainly lie beyond. Therefore, in order to guarantee that the nation's investment in hadron physics facilities and university-based laboratories is fully realized and, furthermore, to ensure that future investments have maximum discovery potential, it is crucial that a flourishing, far-reaching theory program is supported; a program that is balanced optimally amongst complementary needs, which include, *inter alia*, interacting effectively with the nation's experimental effort, capitalizing on the nation's investment in high performance computing, the exploration of novel ideas and new frontiers, and reacting rapidly to the new opportunities that such exploration uncovers.

## 1.3 Endorsements

In addition to considering the material that led to the above recommendations, the QCD and Hadron Physics Town Meeting also considered input from the *Workshop on High Performance Computing* (Computation in Nuclear Physics), Washington DC, July 14-15, 2014, and the *Town Meeting on Education and Innovation*, 6-8 August 2014, NSCL, Michigan State University.

A joint session of the "QCD and Hadron Physics" and "Phases of QCD" Town Meetings voted to endorse the following recommendation of the Workshop on High Performance Computing:

> *Realizing the scientific potential of current and future experiments demands large-scale computations in nuclear theory that exploit the US leadership in high-performance computing. Capitalizing on the pre-exascale systems of 2017 and beyond requires significant new investments in people, advanced software, and complementary capacity computing directed toward nuclear theory.*

along with the following elements of that workshop's request:

> *To this end, we ask the Long-Range Plan to endorse the creation of an NSAC subcommittee to develop a strategic plan for a diverse program of new investments in computational nuclear theory. We expect this program to include*
> - new investments in SciDAC and complementary efforts needed to maximize the impact of the experimental program;
> - development of a multi-disciplinary workforce in computational nuclear theory;
> - deployment of the necessary capacity-computing to fully exploit the nation's leadership–class computers.

The QCD and Hadron Physics Town Meeting also voted to endorse the following conclusions from the Town Meeting on Education and Innovation:

- Education and mentoring of the next generation of nuclear scientists as well as dissemination of research results to a broad audience are integral parts of research.
- Nuclear science is an active and vibrant field with wide applicability to many societal issues. It is critical for the future of the field that the whole community embraces and increases its promotion of nuclear science to students at all stages in their career as well as to the general public.
- Researchers in nuclear physics and nuclear chemistry have been innovative leaders in the full spectrum of activities that serve to educate nuclear scientists as well as other scientists and the general public in becoming informed of the importance of nuclear science. Researchers are encouraged to build on these strengths to address some of the challenges in educating an inclusive community of scientists as well as those on the path to future leadership in nuclear science.
- The interface between basic research in nuclear physics and exciting innovations in applied nuclear science is a particularly vital component that has driven economic development, increased national competitiveness, and attracts students into the field. It is critical that federal funding agencies provide and coordinate funding opportunities for innovative ideas for potential future applications.

## 2. Highlights from the Past Seven Years

The previous Town Meeting on *QCD and Hadronic Physics* took place at Rutgers University in the period 12-14 January 2007. It contributed to a long range plan,[1] which identified a list of overarching questions that define our field:
- What is the internal landscape of the nucleons?
- What does QCD predict for the properties of strongly interacting matter?
- What governs the transition of quarks and gluons into pions and nucleons?
- What is the role of gluons and gluon self-interactions in nucleons and nuclei?
- What determines the key features of QCD, and what is their relation to the nature of gravity and spacetime?

The seven ensuing years have seen considerable progress toward answering these questions, and produced numerous research highlights and discoveries, a selection of which we list chronologically and describe briefly in this Section.

**EMC effect and short-range correlations** – The EMC effect[5] continues to be puzzling. However, important new empirical information has been obtained via JLab inclusive DIS cross section measurements[6] on $^2$H, $^3$He, $^4$He, $^9$Be and $^{12}$C: it does not support previous *A*-dependent or density-dependent fits to the EMC effect and suggests that the nuclear dependence of the quark distributions might depend on the local nuclear environment. Discoveries relating to short-range correlations (SRCs) In nuclei have also been made via triple coincidence (e,e′pN), and inclusive measurements at JLab, which might have important implications for understanding the EMC effect. The JLab Hall-A experiment discovered[7] that neutron-proton pairs are nearly 20 times as prevalent in $^{12}$C as proton-proton pairs and, by inference, neutron-neutron pairs. In a new publication[8] from JLab, results for heavier nuclei – Al, Fe and Pb – in addition to $^{12}$C, show that short-range interactions form high-momentum correlated proton-neutron pairs even in neutron-rich nuclei. This difference between the types of pairs is because of the nature of the strong force and has implications for understanding cold dense nuclear systems such as neutron stars. Analysis of inclusive measurements[9, 10] also establishes quantitatively that the magnitude of the slope of the EMC effect measured in deep inelastic electron scattering in the valence region is linearly correlated with the SRC scale factor obtained from inclusive electron scattering at $x >1$,[11] as shown in Figure 1. New insights into the EMC effect are promised by a number of forthcoming JLab experiments, e.g. a measurement of the polarized EMC effect[12] in $^7$Li and, via comparison between $^{40}$Ca and $^{48}$Ca, a determination of the isospin dependence of the EMC effect, which has implications for interpreting the anomalous NuTeV measurement of the weak mixing angle.[13]

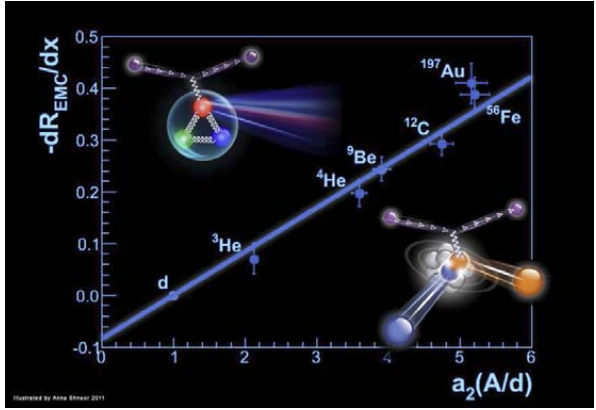

Figure 1 – The EMC slopes versus the SRC scale factors. The uncertainties include both statistical and systematic errors added in quadrature. The fit parameter is the intercept of the line and also the negative of the slope of the line.

**A puzzle surrounding the radius of the proton** – The proton charge radius puzzle appeared recently with reports, in 2010[14] and 2013,[15] of high precision results from muonic hydrogen spectroscopy measurements which are smaller than the values determined from electron scattering experiments[16, 17] and the CODATA compilation by more than $7\sigma$.[18] This puzzle has triggered active theoretical interest, particularly in the context of new physics beyond the SM, and also motivated a range of novel measurements aimed at resolving the mystery, e.g.: a new generation of electron[19, 20] and muon[21] scattering experiments; and new hydrogen spectroscopy measurements.

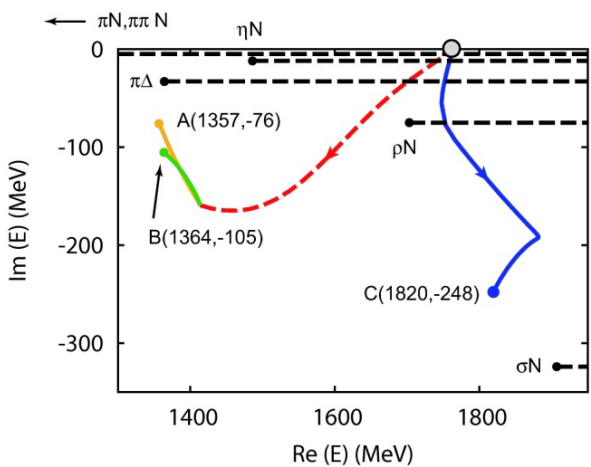

Figure 2 – Owing to complex meson-baryon coupled-channels effects, the "bare" Roper resonance, with mass *1763* MeV, evolves into three distinct spectral features in the $P_{11}$ scattering amplitude as couplings to the meson-baryon continuum are first switched on and then evolved to their physical values. The analysis demonstrates that the lowest two $P_{11}$ resonances found experimentally are actually generated by a single state that may be identified as the radial excitation of the nucleon's "dressed-quark core".

**Dynamical origin of baryon resonances** – Understanding hadron spectroscopy poses many experimental and theoretical challenges. Many excited states are short-lived and close in energy, making it hard to reliably categorize their quantum numbers or to specify their production mechanism. The "Roper resonance", for example, baffled nuclear physicists for almost 50 years. Discovered in 1963, it is just like the proton only 50% heavier. Its mass was the problem: until recently, it could not be explained from QCD by any available theoretical method. That changed following a demonstration that the Roper is the proton's first radial excitation, with its lower-than-expected mass coming from a quark core shielded by a dense cloud of pions and other mesons.[22] This is illustrated in Figure 2. The breakthrough was enabled by new high-quality data obtained at JLab and new analysis tools developed at the Excited Baryon Analysis Center (EBAC), which was located at JLab. This pattern is repeated for several prominent nucleon resonances,[23, 24] although the magnitude of the effect depends strongly on the resonance's quantum numbers.[22,25]

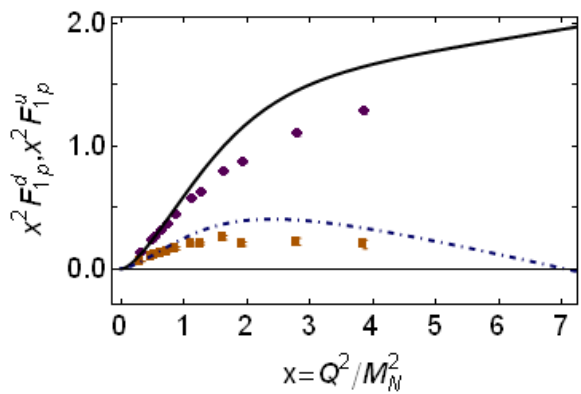

Figure 3 –This figure depicts a flavor separation of the proton's Dirac form factor as a function of $Q^2$. Existing empirical results – *u*-quark (circles) and *d*-quark (squares) – are compared with Faddeev equation calculations,[26, 27] which are distinguished from fits to existing data[28, 29] by the prediction of a zero in the *d*-quark's contribution to the proton's Pauli form factor. The calculations predict that this feature owes to a dynamical interplay between scalar and axial-vector diquark correlations within the proton.[26, 27, 30]

**Flavor separation of the nucleon electromagnetic form factors –** The distribution of charge and magnetization within nucleons is described by the electric and magnetic form factors of these basic constituents of matter. The form factors are empirically accessible in elastic electron-nucleon scattering. Following measurements[31] of the neutron's electric form factor to $Q^2$ = 3.4 GeV$^2$, it became possible to determine the contribution from different quark flavors to the form factors of the neutron and proton.[32, 33] Results for the Dirac form factor are displayed in Figure 3. Whilst a variety of QCD-inspired models can describe existing form-factor data for both the up quark and down quark in the proton at moderate $Q^2$, the calculations diverge dramatically at the larger values of $Q^2$. The Faddeev equation calculation predicts a zero in the down-quark's Dirac form factor, correlates that with the existence and location of a zero in the proton's electric form factor, and connects such qualities the appearance of scalar and axial-vector diquark correlations within the nucleon whose strength and structure is driven by dynamical chiral symmetry breaking (DCSB). The large $Q^2$ domain will be probed with high precision by an array of approved 12 GeV experiments to measure the electromagnetic form factors of both the proton and the neutron.[34, 35, 36, 37, 38, 39] The unprecedented quality and breadth of these experiments will enable us to venture deep inside the nucleon to determine the distributions of charge and currents, and to unravel the flavor structure of the nucleon.

**Parity-violating electron scattering and the structure of the nucleon and nuclei –** Parity-violating (PV) electron scattering has continued to make major advancements as a tool for accessing the strange quark contribution to the electromagnetic structure of the proton, the weak charge of the proton, the neutron radius in heavy nuclei, and for testing the Standard Model of particle physics. Concluding the worldwide

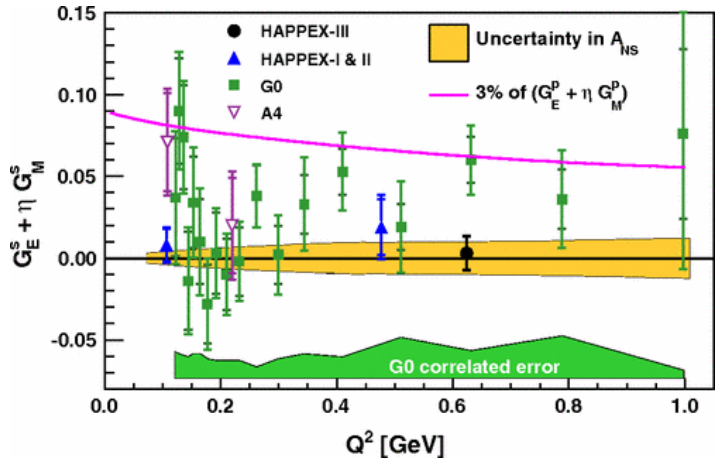

Figure 4 – Results of strange-quark vector form factors for all measurements of forward-angle scattering from the proton. The solid curve (purple) represents a 3% contribution to the comparable linear combination of proton form factors

effort of almost two decades, the newest results[40] from the JLab HAPPEX Collaboration together with the world data show that strange contributions to nucleon form factors are consistent with zero and not more than a few percent of the proton form factors as shown in Figure 4, where all published data on net strangeness contribution $G_E^s + \eta G_M^s$ ($\eta \approx Q^2$) in forward-angle scattering from the proton versus $Q^2$ are presented. The JLab Qweak Collaboration reported the first determination[41] of the weak charge of the proton based on 4% of the total data taken and the result is consistent with the SM prediction. Parity violating electron scattering (PVES) from deuterium in the deep-inelastic scattering (DIS) region at large Bjorken-$x$ is an attractive reaction to search for new physics. The JLab PVDIS collaboration[42] has recently published in Nature a new experimental result that shows the four-Fermi coupling constant of vector (electron) and axial-vector (quark) currents is non-zero, as predicted by the SM.

**Understanding the pion** – Pion properties are intimately connected with DCSB, which explains the origin of more than 98% of the mass of visible matter in the Universe.[43] Enigmatically, owing to the intimate connection between DCSB and the pion, the properties of Nature's lightest hadron provide the most direct access to QCD's momentum-dependent effective quark mass.[44] Consequently, measurement of the electromagnetic form factor of the pion, $F_\pi(Q^2)$, presents an extraordinary opportunity for charting the transition from confinement-dominated physics at large length-scales to the short-distance domain upon which aspects of perturbative QCD become apparent.

Greater urgency is now attached to measurement of $F_\pi(Q^2)$ following recent theoretical progress. Our picture of the pion's valence-quark structure has crystallized with an appreciation that the pion's parton distribution amplitude (PDA) is a broad, concave function, whose dilation is a direct measure of DCSB.[44] Evidence supporting this picture had long been accumulating;[45, 46, 47, 48] and the dilation is now verified by simulations of lattice-QCD.[49] The new picture shows that the pion's valence-quark and -antiquark are more likely than previously thought to have widely differing momenta. Moreover, new methods have enabled direct computation of $F_\pi(Q^2)$ on the entire domain of spacelike momentum transfer, with the prediction that QCD factorization in this exclusive process should be observed for $Q^2 > 8$ GeV$^2$ *but* that in foreseeable experiments the normalization will be fixed by the non-perturbative mass-scale associated with DCSB via the pion's PDA.[50] With the 12 GeV Upgrade, $F_\pi(Q^2)$ can be accurately mapped in Hall C up to momentum transfers of[51] $Q^2 = 6$ GeV$^2$. The large body of additional exclusive charged and neutral pion data expected as part of the 12-GeV program, combined with expected theoretical progress in understanding the reaction mechanism, will potentially extend the kinematic reach to[54] $Q^2 = 9$ GeV$^2$, albeit with slightly lower precision (see Figure 5). These measurements hold great promise: it is possible that they will be the first to sight parton model scaling in an elastic form factor.

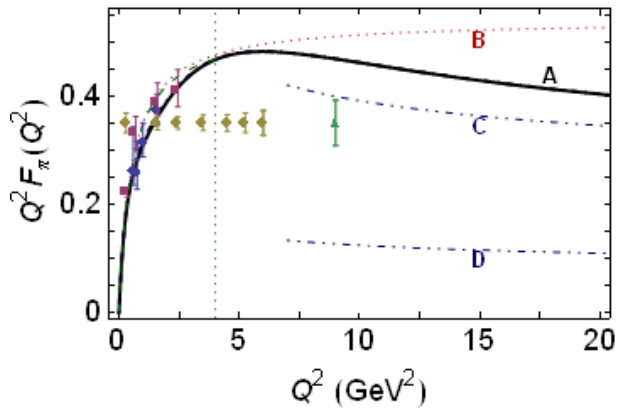

Figure 5 – $Q^2 F_\pi(Q^2)$. Solid curve (A) – Theoretical prediction;[50] dotted curve (B) – monopole form fitted to data;[52] dot-dot-dashed curve (C) – perturbative QCD (pQCD) prediction computed with the modern, dilated pion PDA; and dot-dot-dashed curve (D) – pQCD prediction computed with the asymptotic PDA, which had previously been used to guide expectations for the asymptotic behavior of $Q^2 F_\pi(Q^2)$. The filled-circles and -squares represent existing JLab data[53]; and the filled diamonds and triangle indicate the projected reach and accuracy of forthcoming experiments.[51, 54]

This body of experiment and theory also bear on the controversy that has arisen in connection with the large $Q^2$ results for the $\gamma\gamma^* \to \pi^0$ transition form factor,[55, 56] which is particularly sensitive to the broadening of the pion's PDA and has therefore refocused attention on the need to verify theoretical predictions for the distribution of momentum between the valence quark and antiquark.

**Exotic mesons found on the lattice** – It has long been suspected that QCD might support hybrid mesons; namely, integer spin states with valence gluon content. Such states can possess quantum numbers which are impossible in two-body quantum mechanical systems comprised of a constituent-quark and -antiquark alone; and a primary focus of the GlueX experiment in Hall-D at the 12 GeV JLab facility is the search for these states. Confidence in their existence has recently been boosted by numerical simulations of QCD using lattice methods,[57] which have produced towers of states with exotic quantum numbers in a mass range accessible to the 12 GeV upgrade of JLab, as illustrated in Figure 6. The calculations find that these exotic-quantum number states have a larger gluonic component than normal mesons. Empirical confirmation of the existence of exotic mesons will represent a major advance in our understanding of hadronic matter.

**Resolving the spin and parity of the Λ(1405)** – Modern data has resolved a longstanding puzzle relating to the isospin-zero Λ(1405) baryon, which is an apparently peculiar member of the spectrum. There is still no universal agreement on its character: does it possess a significant dressed-quark core, or is it a resonance, generated purely through meson-baryon coupled channels effects, or a molecular nucleon-kaon bound-state embedded in a Σπ continuum? Whatever its nature, all theoretical attempts to explain its appearance have assumed $J^P=\frac{1}{2}^-$ for the Λ(1405). Until recently, however, empirical confirmation of this assignment was lacking. That has changed with a determination of the spin and parity using photoproduction data from JLab.[58] The reaction $\gamma+p \to K^++\Lambda(1405)$ was analyzed in the decay channel $\Lambda(1405) \to \Sigma^++\pi^-$, where the decay distribution to $\Sigma^+\pi^-$ and the variation of the $\Sigma^+$ polarization direction with respect to that of the Λ(1405) determines the parity. The analysis established that the decays are S-wave, with the $\Sigma^+$ polarized such that the Λ(1405) has spin-parity $J^P=\frac{1}{2}^-$.

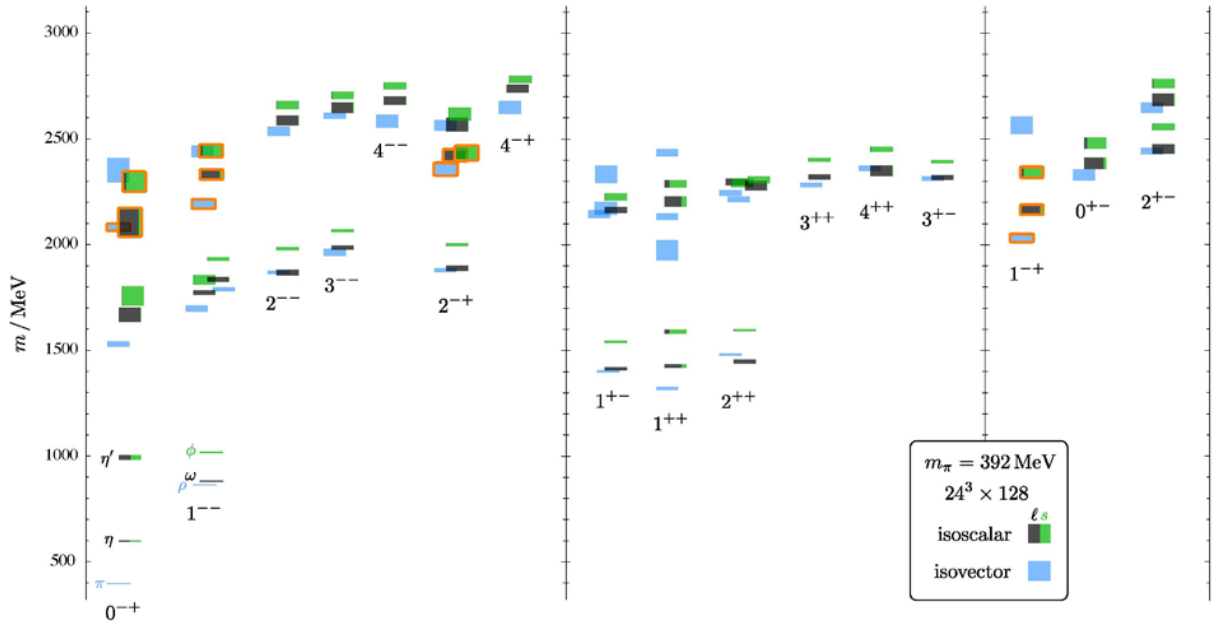

Figure 6 – Isoscalar (green and black) and isovector (blue) meson spectrum obtained with $m_\pi \sim 400$ MeV in a numerical simulation of lattice-regularized QCD. The vertical height of each box indicates the statistical uncertainty on the mass determination. Orange boxes are used to highlight the lowest-lying hybrid states, based on their gluonic field content; and the three rightmost towers of states carry exotic quantum numbers.

**Spin structure of the nucleon and nucleon tomography –** Major progress has been made in addressing the proton spin puzzle since the 2007 LRP: we now have evidence for a gluon spin contribution to the spin of the proton from the RHIC-Spin program, and have witnessed theoretical and experimental advancements in nucleon tomography in the form of the generalized parton distributions (GPDs) and transverse momentum dependent parton distributions (TMDs).

Figure 7 shows the latest STAR[59] and PHENIX[60] data from the 2009 RHIC run for the double-spin asymmetry with longitudinal polarization together with the latest global QCD analysis of polarized parton distributions by de Florian *et al.*[61] These data and the associated analysis have provided evidence that the gluons' spin is preferentially aligned with that of the proton for fractional gluon momenta between 0.05 and 0.2 at the energy scales probed. In addition, experiments at JLab[62, 63, 64] have both yielded first results on the valence quark polarizations at high *x* and mapped the $Q^2$-dependence of various moments of spin structure functions that are connected with higher-twist quark-gluon correlations and, at low-$Q^2$, χEFTs.

These exciting developments plus two decades of accumulated knowledge about the quark contribution to the proton spin and recent theoretical progress[65, 66, 67, 68, 69, 70, 71] further motivate the importance of accessing the parton orbital angular momentum (OAM) contribution to the proton spin. GPDs allow for a determination of the parton OAM contribution to the proton spin through the Ji sum rule.[72] The contribution of quark OAM has been computed in lattice-QCD, with the result that the total angular momentum carried by quarks is small but that of the individual flavors is substantial.[65, 67] While the HERMES and JLab 6 GeV experiments, for the first time, allow for constraints on the *u* and *d* quark orbital angular momentum, upcoming 12-GeV experiments and COMPASS-II with significantly improved precision will provide important tests of predictions obtained with modern QCD theory. Pioneering semi-inclusive deep-inelastic scattering (SIDIS) experiments from HERMES[73], COMPASS[74] and JLab[75] provided the much-needed initial information about TMDs, leading to new phenomena and new dynamics of QCD related to the spin to be discovered in the next decade and beyond. These efforts benefited significantly from the Collins fragmentation functions extracted from the $e^+e^-$ collision data of Belle[76, 77] and Babar.[78]

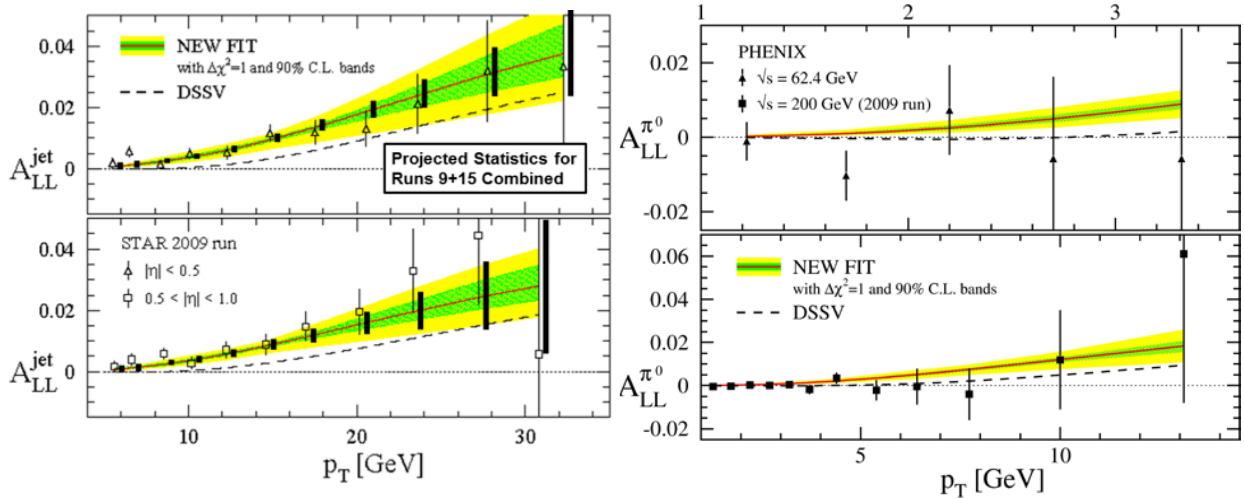

Figure 7 – $A_{LL}^{jet}$. Latest STAR[59] and PHENIX[60] data for the double-spin asymmetry in jet production for two rapidity ranges compared to the results of the new[61] (solid line) and original[79] (dashed) analyses by de Florian et al. The inner and outer bands correspond to $\Delta\chi^2=1$ unit and 90% C.L., respectively. The thick black bands in the left panel indicate the projected statistical precision for inclusive jets in 200 GeV $p+p$ collisions based on combined data from the 2009 and 2015 RHIC runs.

**Compton scattering and nucleon polarizabilities** – Polarizabilities parametrize the deformation of nucleons in electromagnetic fields and are thus benchmarks for our understanding of hadron structure. They are also crucial to developing understanding of the neutron-proton mass difference[80] and the proton's charge radius.[81] Chiral effective field theory (χEFT) facilitates extraction of these hadron-structure parameters from Compton cross sections.[82] Recent efforts to determine nucleon polarizabilities have relied on a synergistic blend of theory and experiment, e.g., in a collaboration to obtain and interpret γ-deuteron scattering data from MAX IV in Sweden, which reduced uncertainties in the neutron polarizabilities by one-third.[83] The experimental effort is moving to HIGS and Mainz and increasingly focusing on spin polarizabilities, thereby complementing investigations of nucleon spin structure at JLab and elsewhere. A pioneering measurement of doubly-polarized Compton-scattering was recently performed at Mainz, enabling all four proton spin polarizabilities to be obtained for the first time.[84]

## 3. Physics of the Future

### 3a. Hadron theory

A key fascination of QCD is that it is possibly a nonperturbatively well-defined quantum field theory.[85] If so, then it would be unique within the Standard Model. Additionally, there is no confirmed breakdown of QCD over an enormous energy domain: *0* GeV < *E* < *8* TeV; and results from the LHC have led to a resurgence of interest in the possibility that any extension of the Standard Model will be based on the paradigm established by QCD. These features and possibilities lend urgency to the problem of solving QCD, and hence to modern programs in hadron physics.

Central to solving QCD is an elucidation of the nature of confinement and its connection with DCSB.[1, 43] It is now widely accepted that in the presence of light-quarks, confinement cannot be associated with the flux tube that is drawn between two static color sources. Instead, confinement appears to be a dynamical process. Contemporary theory predicts that *both quarks*[1, 43] *and gluons*[86, 87] acquire running mass distributions in QCD, which are large at infrared momenta. The generation of these masses leads to the emergence of a length-scale $\sigma \approx 1/2\Lambda_{QCD} \approx 0.5$fm, whose existence and magnitude is evident in all existing studies of dressed-gluon and -quark propagators and which characterizes a dramatic change in the analytic structure of the these propagators.[88, 89] In models based on such features, once a gluon or quark is produced, it begins to propagate in spacetime; but after each "step" of length $\sigma$, on average, an interaction occurs so that the parton loses its identity, sharing it with others. Finally a cloud of partons is produced, which coalesces into color-singlet final states. Such pictures of parton propagation, hadronization and confinement can be tested at the upgraded JLab facility and a future EIC.

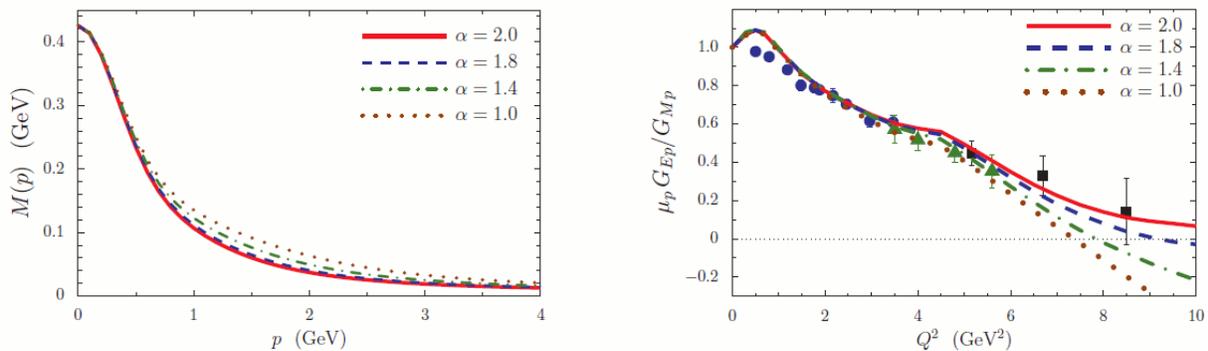

Figure 8 – *Left panel*: QCD's quarks do not have a fixed mass but instead possess a mass distribution, *M(p)*. The shape of this distribution is predicted to explain the origin of the bulk of visible mass in the Universe. At present, neither experiment nor theory can distinguish between the various mass distributions that are illustrated in this figure. The parameter $\alpha$ is a theoretical device used to model just how much of the visible mass in the universe is generated by strong interaction dynamics. *Right panel*. Computed response of the ratio $G_E(Q^2)/G_M(Q^2)$ as a function of $\alpha$.[90] Evidently, this ratio is a sensitive measure of the dressed-quark's mass distribution and therefore experiments planned for the upgraded JLab facility will be able to tightly constrain the nature of that distribution.

As described in connection with Figure 5, the last seven years have seen material progress in understanding the pion and exposing the enormous impact of DCSB. Crucial to these successes is an emerging ability to compute Poincaré covariant ground-state hadron wave functions. This enables theory to expose the connection between QCD's emergent phenomena and measurable quantities, as explained in connection with Figure 3. Another striking illustration is provided by calculations of the ratio of proton electric and magnetic form factors, which were an empirical highlight in the 2007 Long Range Plan.[1] As illustrated in Figure 8, measurements of the ratio $G_E(Q^2)/G_M(Q^2)$ appear to be a keen probe of the mass distribution associated with dressed-quarks within a hadron.

Given the challenges posed by non-perturbative QCD, it is insufficient to study hadron ground-states alone. Many novel perspectives and additional insights are provided by nucleon-to-resonance transition form factors, whose behavior at large momentum transfers can reveal much about the long-range behavior of the interactions between quarks and gluons.[91] Indeed, in analogy with exotic and hybrid states, the properties of nucleon resonances are more sensitive to long-range effects in QCD than are those of hadron ground states. The lightest baryon resonances are the *Δ(1232)*-states; and despite possessing a width of 120 MeV, these states are well isolated from other nucleon excitations. Hence the *γ+N→Δ* transition form factors have long been used to probe strong interaction dynamics. They excite keen interest because of their use in probing, *inter alia*, the relevance of perturbative QCD to processes involving moderate momentum transfers;[92, 93, 94] shape deformation of hadrons;[95] and, of course, the role that resonance electroproduction experiments can play in exposing non-perturbative features of QCD.[91] Using CLAS at JLab, precise data on the dominant *γ+N→Δ* magnetic transition now reaches to $Q^2$ = 8 GeV$^2$; an eventuality that poses both great opportunities and challenges for QCD theory, some of which have recently been met, as illustrated in Figure 9.

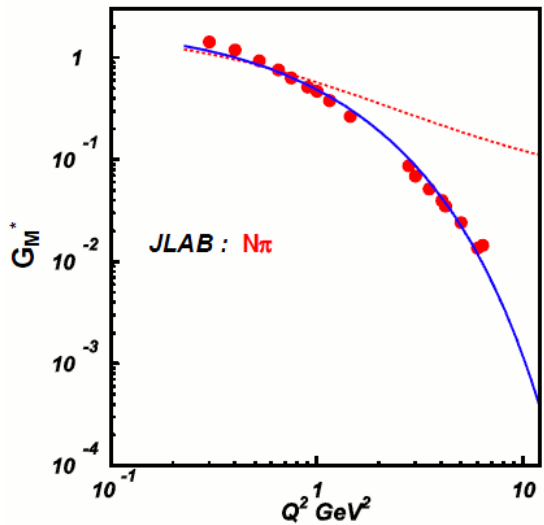

Figure 9 – Comparison between CLAS data[23] on the magnetic *γ+N→Δ* transition form factor and a recent theoretical prediction.[27] The dashed curve shows the result that would be obtained if the interaction between quarks in QCD were momentum-independent. The solid curve is obtained with precisely the same QCD-based formulation as was employed for the nucleon elastic form factors in Figure 3 and Figure 8. The experiment-theory comparison establishes that experiments are sensitive to the momentum dependence of the running couplings and masses in QCD; and the theoretical unification of *N* and *Δ* properties highlights the material progress that has been made in constraining the long-range behavior of these fundamental quantities.

With the growing ability to calculate Poincaré-covariant hadron wave functions it is becoming possible to predict and understand the distribution of partonic matter within hadrons. This enables the impact of intra-hadron correlations on parton distributions to be exposed. The valence-quark domain is of particular interest for many reasons. For example, valence-quark structure is definitive of a hadron – it's how one tells a proton from a neutron and it expresses every one of a hadron's Poincaré-invariant properties. Moreover, although parton distributions all vanish at Bjorken-*x=1*, the ratio of any two need not; and the value of such ratios is invariant under QCD evolution.[96] The ratios are therefore a scale-invariant, non-perturbative feature of QCD, which provide a sharp discriminator between frameworks that claim to explain hadron structure. In this connection, it has recently been shown[97] that correlations between dressed-quarks within the nucleon have a very significant impact on both unpolarized and polarized valence-quark distribution functions. Hence experiments planned at JLab-12,[98, 99, 100, 101, 102, 103] and elsewhere, aiming to extract the ratio $d_v/u_v$ and nucleon longitudinal spin asymmetries at large *x*, promise to add considerably to our knowledge of nucleon structure in the foreseeable future.

Lattice-regularized Quantum Chromodynamics (lQCD) is a powerful numerical method that enables key properties of the theory of strong interactions to be computed from first principles in the strong-coupling regime. Thus, lQCD calculations can determine the bound states of the theory, and describe how the quarks and gluons of QCD give rise to the observed protons, neutrons, pions, and the other hadrons. They can determine how charge, current, and matter are distributed within a hadron, and contribute to building a three-dimensional picture of the proton. The emergence of the nuclear force from QCD can be investigated, leading to a refinement of chiral nuclear forces; and first-principle calculations of the structure and reactions of light nuclei can be performed.

Lattice calculations are expected to play a key role in complementing, supporting and fully capitalizing on the current and future DOE experimental nuclear physics programs. Lattice QCD calculations will predict the spectrum and properties of so-called exotic mesons, states in which gluonic degrees of freedom might be manifestly exposed and whose discovery is a primary aim of the GlueX experiment at the 12 GeV upgrade of Jefferson Laboratory. Calculations of nucleon form factors, generalized parton distributions, and transverse-momentum-dependent distributions will aid in drawing a more complete three-dimensional tomography of the nucleon than the experimental programs at JLab and at RHIC-spin can alone provide.

The past five years have witnessed numerous achievements that have advanced our understanding of QCD. Lattice calculations of the spectrum of low-lying isovector mesons[57, 104, 105, 106] have suggested the presence of hybrid mesons, those in which the gluonic degrees of freedom are manifest, in an energy regime accessible to GlueX at JLab-12 (see Figure 6). Lattice calculations have been performed of the moments of GPDs,[107, 108, 109] delineating the contribution of quark orbital angular momentum to nucleon spin, and TMDs,[110, 111] key elements in the experimental programs both of RHIC-spin and of JLab. Finally, our ability to derive an *ab initio* understanding of the interactions between nucleons has been demonstrated.[112, 113]

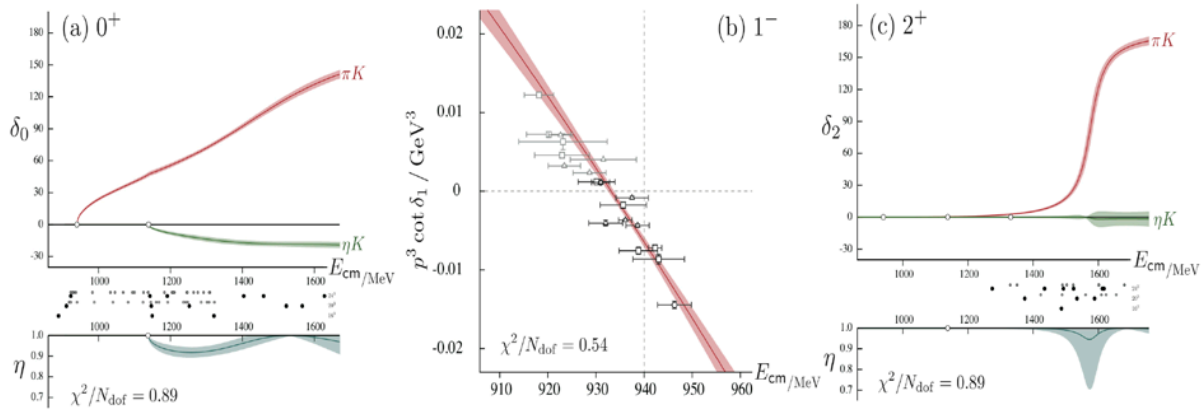

Figure 10 – (a) *Upper panel*: $J^P = 0^+$ momentum-dependent phase shifts in the coupled $\pi K$, $\eta K$ system; and *lower panel*: the inelasticity. (b) $J^P = 1^-$ phase shifts around the $K\pi$ threshold. (c) $J^P = 2^+$ amplitudes and inelasticity. (Lattice-QCD calculations of scattering.[114])

The advent of leadership-class exascale computing, and new algorithmic and theoretical ideas over the next five years provide an unprecedented opportunity for lQCD to help advance our understanding of nuclear physics still further. Calculations of the spectrum of QCD, with scattering amplitudes for inelastic channels faithfully included, are now within reach[114, 115] (see Figure 10) and will proceed by capitalizing on experimental advances in spectroscopy; and new methods will be exploited to explore the electromagnetic properties of nucleon resonances.[116] Moreover, in a potentially significant recent development, it has been suggested[117] that light-cone correlation functions may be computed directly from boosted Euclidean space correlation functions, thereby overcoming the historical difficulty associated with access to only the lowest few moments of these functions in conventional lattice analyses.

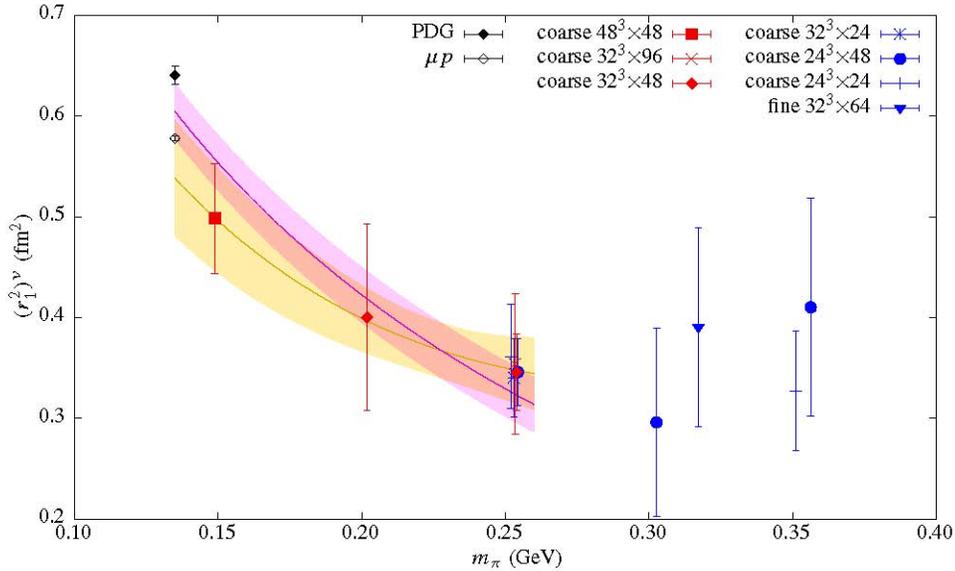

Figure 11 – Isovector Dirac charge radius from a lattice calculation at near-to-physical quark masses:[118] the bands are two possible chiral extrapolations of the lattice calculations. The experimental points are those of the CODATA compilation[18] used by the PDG, and the result obtained from muonic hydrogen.[14, 15]

Precise calculations of hadron structure at physical quark masses, exemplified by near-to-physical calculations of electromagnetic form factors,[118] illustrated in Figure 11, and the exploration of novel proposals to compute some key measures of hadron structure will be important to the imaging of hadrons at a future Electron-Ion Collider. However, full exploitation of these opportunities requires investment in algorithm and software development, and capacity computing.

Effective field theories (EFTs) are powerful tools when tackling problems with a natural separation of energy scales. In such instances they provide a systematic expansion of measurable quantities in terms of parameters that may be determined from experiment, or computed theoretically when reliable methods are available. EFTs are particularly useful in QCD, where the relevant degrees of freedom range from quarks and gluons at high energy to hadrons and nuclei at lower scales. Important examples in nuclear physics are provided by chiral EFTs (χEFTs) and the schemes used to extrapolate lQCD results to the continuum limit. Recent progress in the use of chiral EFTs is exemplified by analyses of Compton scattering and nucleon polarizabilities, highlighted in Sec. 2. It also includes the analysis of lQCD results in order to compute the *I=2 ππ S*-wave scattering phase shift[119] and a marriage of EFT with dispersion relations so as to aid in understanding the contribution of hadronic light-by-light scattering to the muon's anomalous magnetic moment, $(g-2)_\mu$.[120] Valuable insights into the nature of the baryon spectrum have also been drawn from the combined use of the $1/N_c$ expansion, heavy-baryon χEFT and lQCD.[121]

Factorization of high-energy cross-sections in QCD enables extraction of information about both the distribution of gluons and quarks within hadrons and nuclei and correlations between them. Significant advances have recently been made in: the precision of perturbative calculations of partonic scattering – *sharpening the probe*; the extraction of the hadronic matrix elements – *better knowledge of structure*; and developing factorization for new physical observables – *new probes and new structures*. Factorization is consistent with the concepts of effective field theory (EFT), since non-perturbative physics associated with the observed energetic hadrons is approximated by well-defined and universal hadronic matrix elements. Soft-collinear effective theory (SCET)[122, 123, 124, 125] is designed for treating processes with energetic hadrons, and is a natural EFT for investigating QCD dynamics at colliders.

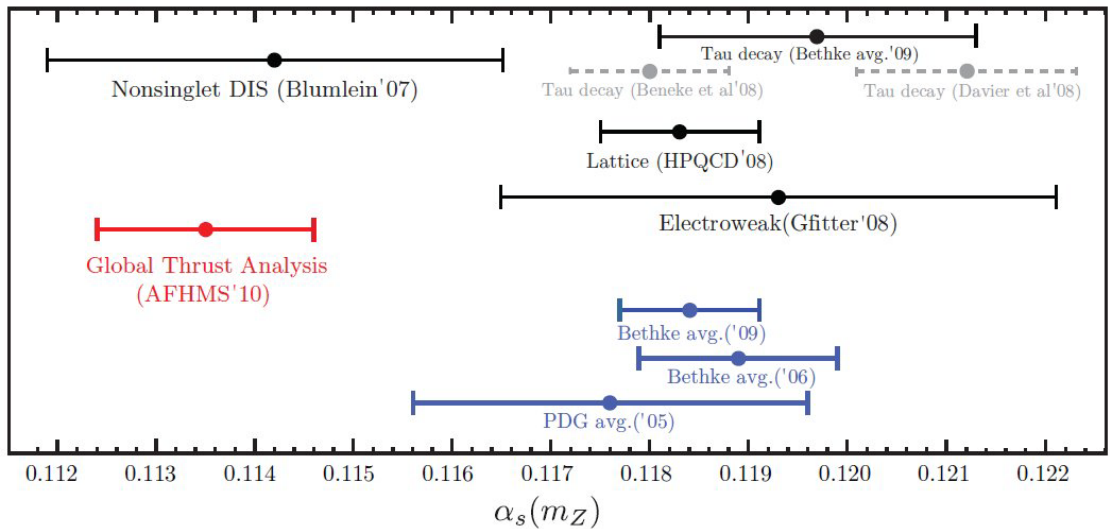

Figure 12 – Selected determinations of $\alpha_S$ ($m_Z$) defined in the MS-bar scheme compared with that obtained using SCET.[159] The high energy extractions from thrust (SCET) and DIS[126] appear to be smaller than the low energy extractions from lattice-QCD[127] and τ-decays,[128, 129, 130] and the global averages[131, 132, 133] that those estimates influence.

SCET techniques have successfully been applied to QCD calculations, especially in organizing high-order corrections and resummations of large logarithms for processes involving multiple well-separated momentum scales.[134, 135, 136, 137, 138, 139, 140, 141] SCET also plays an increasing role in identifying new and potentially factorizable observables, which enables improved tests of hard QCD dynamics.[142, 143, 144, 145, 146, 147, 148] The precision of the *probe* at colliders – calculable partonic scattering – is characterized by a perturbative expansion in powers of the strong coupling constant, $\alpha_S$, and the resummation of large perturbative logarithms. The state-of-the-art for computation of $e^+e^- \to$ jets is now $O(\alpha_S^3)$ – N³LO accuracy;[149, 150] NLO partonic cross-sections for lepton-hadron and hadron-hadron collisions are now available for observables with more jets;[151, 152] and the NNLO frontier has been extended to many critically important partonic processes: $gg \to gg$, $pp \to tt$, $gg \to H$+jet, etc.[153, 154, 155, 156] Theoretical uncertainty for inclusive jet production is now just a few percent, an unprecedented accuracy for a QCD calculation.[157] With the aid of SCET, the resummation of next-to-next-to-next-to-leading-logarithms

($N^3$LL) has been achieved for the thrust distribution in $e^+e^-$ collisions, which has enabled a much more accurate extraction of $α_S$,[158,159] (see Figure 12).

For various event-shape observables relevant for a future EIC, results at $N^2$LL order are now available,[146, 160, 161] with results at $N^3$LL order expected soon.[162] Notably, too, with the clear separation of jet energy and temperature of the medium, SCET techniques are now applied to investigate jet quenching and jet structures in a hot, dense quark-gluon medium, marking a clear cross-fertilization between "cold" and "hot" QCD research in nuclear physics.[163, 164, 165, 166]

Regarding structure, the QCD evolution kernel for parton distribution functions (PDFs) is now available to $O(α_S^3)$.[167] With the tremendous amount of data available from JLab, RHIC, and the LHC, and improved calculations of partonic cross-sections, the accuracy of extracted PDFs has been steadily improving.[168] Now, with less than 30% uncertainty from PDFs, factorized-QCD predictions for production of inclusive jets in hadronic collisions agree with data over many orders-of-magnitude and up-to 2 TeV in jet transverse energy.[169]

With JLab12 being ideal for the valence-quark region and a future EIC providing unprecedented access to gluon and sea-quark distributions, SIDIS offers a unique opportunity to probe the confined transverse motion of partons inside a colliding nucleon or nucleus. SIDIS naturally possesses a two-scale event structure. The large virtuality of the exchanged vector boson, $Q$, localizes the probe whilst the transverse momentum of the produced hadron, $p_T \ll Q$, in the frame where the exchanged boson and the nucleon or nucleus collide head-on, is sensitive to the momentum scale of the confined motion of gluons and quarks. Theory has recently made significant progress toward factorizing the measured SIDIS cross-sections, such that they are expressed systematically in terms of transverse momentum dependent PDFs (TMDs). QCD predicts how TMDs evolve with $Q$;[140, 170] and TMDs extracted from SIDIS could be compared with those extracted from $W$-production at RHIC to test such evolution. Notably, however, with very limited SIDIS data available, various groups have "predicted" very different behavior of the TMDs, despite starting with the same evolution equations.[171, 172, 173, 174, 175] More theoretical work and new data are urgently needed in order to resolve these conflicts. Since factorization in terms of TMDs is more likely to be broken in hadronic collisions,[176, 177] JLab12 and a future EIC are essential.

Notwithstanding the progress that has been made in QCD theory, the complexity and diversity of strong interaction phenomena ensure that phenomenology, and the construction and use of QCD-inspired models remain an essential part of the hadron theory effort. At the very least, such methods enable connections to be drawn between the results of *ab initio* computations and experimental data. Even more importantly, perhaps, they enable the rapid development of insights and intuition regarding complex systems and reactions, thus providing the information that is very often necessary for grasping the key elements of a new discovery and planning the next sensible step.

As emphasized by the success of EBAC at JLab and kindred analysis efforts,[178, 179] phenomenology is crucial in hadron spectroscopy, which is not *bump hunting* but the search for poles in the complex energy plane. Analyses of data must incorporate *S*-matrix constraints and state-of-the-art knowledge of reaction dynamics; and in this endeavor, a synergy of experiment, theory and phenomenology is essential. The interpretation of experiments requires QCD calculations in the continuum and on the lattice, and QCD-inspired models of confinement. Critically, the search must acknowledge and understand that most of the objects being sought are unstable: decays are an essential part of a hadron's brief existence, and the hadron spectrum will only be charted once the impact of the final states is thoroughly understood, as highlighted in Figure 2. EBAC has evolved into JPAC, the JLab Physics Analysis Center. This is a multi-institution collaboration charged with building a sophisticated analysis framework that: captures the significant reaction mechanisms; incorporates crossing symmetry; expresses unitarity and the impact of final-state interactions; and uses the best model-dependent and model-independent amplitude analyses.

Constituent-quark models of hadrons have long played an important role in explaining and ordering the hadron spectrum. They are of continuing value because their simplicity guarantees both a wide reach and a capacity to shed light on phenomena that lay beyond the reach of more sophisticated approaches. A recent addition to this resourceful toolkit is holographic QCD,[180] which is an extra-dimensional approach to modeling hadrons. In holographic models, the extra spatial dimension creates a waveguide for fields, and the discrete towers of modes propagating in that waveguide are interpreted as hadronic resonances. These models are motivated by the AdS/CFT correspondence, which is a duality that relates theories in different numbers of spatial dimensions: they apply techniques, originally developed in string theory, to hadron physics. Holographic models provide novel perspectives on the forces which produce color confinement, predictions for meson and baryon spectroscopy, and phenomenologically successful forms for the light-front wavefunctions that underlie much of hadron dynamics – including form factors, distribution amplitudes, structure functions, GPDs, etc.

Insightful phenomenology is also currently playing a key role in scanning the horizon made visible by the rapidly expanding body of empirical information on GPDs and TMDs, which was highlighted in Sec. 2. These quantities are united via the concept of gluon and quark Wigner distributions.[181, 182, 183, 184] Computations of GPDs and TMDs within frameworks that possess a direct connection with QCD are in their infancy, so progress is currently being made primarily through the development of efficacious models, and fits and parametrizations of GPDs and TMDs.[185, 186]

### 3b. Hadron structure at short distances

The ongoing and planned experiments will continue to explore the internal landscape of the nucleon, including its spin contributions. In particular, these investigations aim to extend the one-dimensional parton picture to multi-dimensional tomography of partons.

**JLab 12 GeV Upgrades** – The 12 GeV Upgrade will provide the important combination of high beam intensity and reach in $Q^2$ to allow us to map out the quark distributions, both the polarized and unpolarized ones in the valence region, through the measurements of inclusive and semi-inclusive (spin) structure functions at large x with unprecedented precision. With these measurements, we will be able to map out the flavor dependence of the polarized valence and sea quark distributions and significantly improve the extraction of the polarized gluon distribution at large *x*. For unpolarized quark distributions, there are complementary and independent methods to tackle the long-standing issue on the ratio of down to up quarks in the proton, *d(x)/u(x)*, whose large–*x* behavior is intimately related to the fact that the proton and neutron, and not the Δ, are the building blocks of atomic nuclei. These measurements will have a profound impact on our understanding of the structure of the proton and neutron. They will also provide crucial inputs for calculating cross sections for hard processes that will be used at high-energy hadron-hadron colliders such as the LHC for tests of the Standard Model (SM) or for searches for new physics beyond SM.

As mentioned previously (Sec. 2), the pioneering efforts of HERMES, COMPASS, Belle and Babar, together with the 6-GeV Jefferson Lab, have demonstrated the feasibility of studying Transverse Momentum Distributions (TMDs) as well as the Generalized Parton Distributions (GPDs). The extended kinematic range and new experimental hardware associated with the Jefferson Lab 12 GeV Upgrade will provide access to these fundamental underlying distributions and reveal new aspects of nucleon structure, the three-dimensional tomography.

Deeply Virtual Compton Scattering (DVCS) and Deeply Virtual Meson Production (DVMP) are the most powerful processes for providing the necessary observables to perform the spatial tomography of the nucleon for each constituent flavor. The increased energy of the electron beam to 12 GeV offers not only a reach of momentum transfer allowing for the leading order GPD formalism to be applicable, but also provides the highest polarized luminosity for precision measurements of key polarization observables crucial in these studies. A suite of approved DVCS and DVMP experiments planned in Hall B with CLAS 12, Hall A, and Hall C will provide the necessary high precision data for different channels and reactions over a wide kinematical range to access the GPDs. As a direct consequence of the space-momentum correlation there is a way to reach the contribution the orbital angular momentum of quarks makes to the nucleon's spin through the Ji sum rule.[72]

Previous experiments on semi-inclusive deep inelastic lepton scattering (SIDIS) have offered a first glimpse of the effects of transverse motion of quarks, i.e., the TMDs, and the way this is correlated with either their own spin or that of the nucleon. The JLab 12-GeV era can move this field to a new level of sophistication thanks to the extraordinary statistical accuracy achievable and the extended kinematic reach. Each hall brings an essential element to the SIDIS campaign: Hall A with Super-BigBite and SoLID, Hall B with CLAS 12, and precision SIDIS experiments in Hall C, will allow a far more refined determination of the TMDs, thereby enabling a high-resolution momentum-tomography of nucleon. Moreover, precision measurements of spin and azimuthal asymmetries in semi-inclusive pion and kaon production from unpolarized and longitudinally and transversely polarized targets will allow extraction of the spin and flavor dependence of quark TMDs in the valence region;[187, 188, 189, 190, 191, 192, 193, 194, 195, 196] and measurements in the wide range $Q^2 \in (1,6)$ GeV$^2$ will allow studies of the $Q^2$-dependence of TMDs and, in particular, the Sivers TMD,[195] which is predicted to change noticeably under evolution.[197, 198]

The tensor charge is an important intrinsic property of the nucleon, similar to its axial charge or magnetic moment, and corresponds to the lowest moment in $x_B$ of the transversity distribution function $h_1(x)$. It offers a benchmark test for the most modern Lattice QCD calculations, predictions based on continuum QCD, and phenomenology. This distribution is accessible in SIDIS, through the well-known Collins' effect, by using transversely polarized targets. It will be measured in Hall B using CLAS12 and in Hall A using SoLID, where the unique feature of large acceptance and high luminosity will be utilized simultaneously with both polarized neutron ($^3$He) and proton targets. The flavor separated tensor charges of the nucleon will be determined with much improved precision in the 12 GeV era as compared to previous constraints. Figure 13 shows the projected JLab 12-GeV SoLID determination of the *u-* and *d-* quark contributions to the proton's tensor charge (black points) together with model-dependent extractions of these quantities based on existing data. While many assumptions were involved in such extractions, the comparison shows the power of high precision data over multi-dimensional kinematic variables. Also shown are predictions from lattice QCD, Dyson-Schwinger Equations, and various models.

Plans[212, 213] are in place at JLab to construct a PbWO$_4$ electromagnetic calorimeter for neutral particle detection. The combination of neutral-particle detection and a high-resolution magnetic spectrometer offers unique scientific capabilities to push the energy scale for hard exclusive and semi-inclusive processes requiring precision and high luminosity at 12 GeV CEBAF. Such a combination enables precision measurements of DVCS cross section at different beam energies to extract the real part of the Compton form factor without any assumptions. It further makes possible measurements of the basic semi-inclusive neutral-pion cross section in a kinematical region where the QCD factorization scheme is expected to hold, crucial to validate the foundation of this cornerstone of 3D transverse momentum imaging.

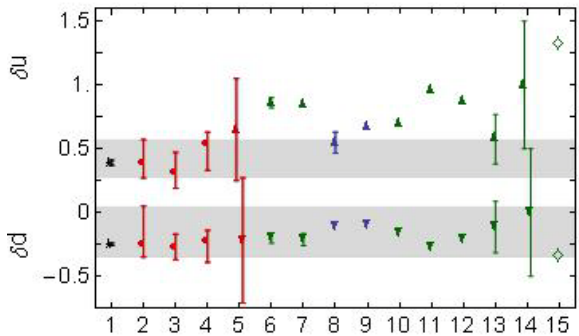

Figure 13 – Compilation of our current knowledge about the *u*- and *d*-quark tensor charges determined from analyses[199, 200, 201] of existing data (shown as points 2-5), the projected results from the JLab SoLID program within the same model[202] (point 1), together with predictions based on lattice QCD[203, 204] (points 6, and 7), Dyson-Schwinger equations[205, 206] (points 8 and 9), and from various models[207, 208, 209, 210, 211] (points 10 - 15). The model-dependent uncertainty in the latest extraction[111] is shown as a grey band.

**RHIC-Spin Program** – RHIC addresses key open questions on the nucleon spin in various complementary ways. It makes use of the unique versatility of the accelerator, which enables collisions of longitudinally and transversely polarized protons relative to their momenta readily available at two main experiments.

The latest data from the 2009 RHIC run with longitudinal polarization have, for the first time, provided evidence that gluons have a preferential alignment of their spins with the proton's spin. This is a milestone for the field, offering new insights into the proton spin decomposition and the nature of the strong force fields inside a proton. The impact of this data is shown in Figure 7 and compared with near-term prospects from the analysis of RHIC top-energy data and the run in 2015. Detector upgrades during the next few years will allow extending this sensitivity to gluons with fractional momenta smaller than 1%.

At RHIC one uses a powerful technique based on the violation of parity in weak interactions. The $W^{\pm}$ bosons naturally select left quark handedness and right antiquark handedness and hence are ideal probes of nucleon helicity structure. Data with longitudinal polarizations of the proton beams from RHIC have now reached the precision needed to obtain sensitive constraints on the helicity distributions of the light quarks and anti-quarks, and a significant further improvement in precision is anticipated when ongoing analyses are finalized.

Among the quantities of particular interest are unique parton distribution functions that can be accessed only in hard-scattering reactions involving transversely polarized protons, including the quark transversity distributions that are linked to the nucleon tensor charge discussed above, and the Sivers functions. RHIC experiments will provide complementary methods to tackle in the quark transversity distributions through various observables. Together with other proposed measurements with unique upgrades, they will help identify the physics mechanism that underlies the large value of $A_N$ in hadronic collisions, a longstanding puzzle.[214]

In particular, the "Sivers functions" express correlations between a parton's transverse momentum inside the proton and the proton spin vector. As such the Sivers' functions contain information on orbital motion of partons in the proton. It was found that the Sivers functions are not universal in hard-scattering reactions. This by itself is nothing spectacular; however, closer theoretical studies have shown that the non-universality has a clear physical origin that may broadly be described as a re-scattering of the struck parton in the color field of the remnant of the polarized proton. Depending on the process, the associated color Lorentz forces will act in different ways on the parton. In semi-inclusive deep-inelastic scattering (SIDIS) processes, the final-state interaction between the struck parton and the nucleon remnant is attractive. In contrast, for Drell-Yan processes, the interactions occur in the initial-state interaction and are repulsive. As a result, the Sivers' functions have opposite signs in SIDIS and Drell-Yan processes.[215, 216, 217] This is a fundamental prediction about the nature of QCD color interactions, directly rooted in the quantum nature of the interactions. There have been strong efforts in the hadron physics community to observe this change of sign. Specifically, experiments at RHIC plan to measure the Sivers single spin asymmetries in Drell-Yan type processes in hadronic collisions, as do the COMPASS experiment at CERN and the polarized Drell-Yan experiments at Fermilab. The complete RHIC-Spin program, including a discussion of future opportunities, is described elsewhere.[218] Proposals have been developed for a high-luminosity polarized Drell-Yan program at the Fermilab Main Injector with both polarized beams and targets. Such a program would leverage the investment of the U.S. Nuclear Physics community in the SeaQuest experiment and the existing facilities at Fermilab.

### 3c. Hadron structure at long distances

**Form factors of nucleons** – Elastic form factors are of fundamental interest and widespread value because they express the distributions of charge, magnetization and spin within the non-pointlike hadrons that QCD is supposed to generate. Their measured forms are therefore a benchmark test for phenomenology and theory within QCD; and also crucial inputs to calculations and experiments in both atomic physics and studies of nuclear structure.

The 6 GeV era at JLab produced dramatic improvements in our understanding of nucleon form factors.[219, 220, 221, 222] These successes enabled a flavor separation of the nucleon form factors, the importance of which was highlighted in Sec. 2. One of the major thrusts of the 12 GeV program at JLab is obtaining new high-quality data on nucleon form factors and, therewith, flavor separations that extend over a much greater domain. These experiments will reveal whether $G_{En}$ and $G_{Ep}$ become negative and whether the electric form factor of the charge-neutral neutron actually becomes larger than that of the charge-one proton. In addition, ongoing and planned experiments are expected to yield a better understanding of those radiative corrections which are currently thought to explain the difference between Rosenbluth and polarization transfer extractions of $G_{Ep}(Q^2)/G_{Mp}(Q^2)$. Furthermore, it is anticipated that improved data will resolve low-$Q^2$ mismatches between different electron scattering measurements,[16, 17, 223, 224] and also the greater puzzle of the discrepancy between the proton radii obtained in atomic physics and elastic electron scattering measurements, highlighted in Sec. 2.

New experiments with neutrino beams at Fermilab will use neutral-current elastic neutrino-proton scattering to measure the full proton axial form factor for the first time. The MicroBooNE experiment[225] in particular is ideally suited to observe the low-$Q^2$ single-proton tracks that are the signature of neutral-current elastic scattering. Combined with charged-current data measured at the same $Q^2$ values, the strangeness contribution to the proton axial form factor can be isolated. It will be used to determine the total strange quark contribution to the proton spin, $\Delta S \equiv \Delta s + \Delta \bar{s}$, in a method that is independent of the measurements planned using polarized DIS at a future EIC.[226] Knowledge of the strangeness contribution to the axial form factor is critical to our studies of nucleon structure and vital to searches for heavy dark matter particles.[227] MicroBooNE will begin collecting data in the first half of 2015.

**Meson form factors** – The form factors of pions and kaons are of special interest owing to the dichotomous nature of these mesons as both bound-states of strongly-dressed constituents and the *pseudo*-Goldstone modes arising through dynamical chiral symmetry breaking (DCSB) in QCD. Recent years have seen dramatically improved precision in measurements of the pion's elastic form factor, $F_\pi(Q^2)$,[53] and birth of a controversy concerning the behavior of the $\gamma^*\gamma\pi^0$ transition form factor, $F_{\gamma^*\gamma\pi}(Q^2)$.[228, 229] The new and improved data have driven renewed theoretical activity, which has focused on the pointwise form of light meson parton distribution amplitudes and their connection with DCSB, as highlighted in Sec 2; and the contribution of transversely polarized photons to meson cross-sections.

Experimentally, pion elastic form factor measurements at JLab are made indirectly, using exclusive pion electroproduction, $p(e,e'\pi^+)n$, to gain access to the proton's "pion cloud". This approach is reliable in forward kinematics.[230] Analogously, in order to extract information on the kaon's elastic form factor it might be feasible[231, 232] to sample the proton's "kaon cloud" via $p(e,e' K^+)\Lambda$. In this instance, JLab at 12 GeV is essential for the measurements at low $t$ that would allow for a clean interpretation of the kaon pole contribution. This data could allow for valuable comparisons between the $Q^2$ dependence and magnitude of the $\pi^+$ and $K^+$ form factors.[233]

**Resonance transition form factors** – The excitation of nucleon resonances was a core component of the 6 GeV program at JLab and it will continue with operations at 12 GeV. This component studies the formation of excited nucleon states and their emergence from the interactions between dressed quarks in QCD. Using the CLAS detector, the first high precision photo- and electroproduction data have become available[24, 25, 234] and, as highlighted in Sec. 2, this data led to a new wave of significant developments in reaction theory, and in the phenomenology and theory of QCD. The large number of nucleon-to-resonance transition form factors, with their diverse array of features, ensures that entirely new windows on hadron structure are opened by studying the $Q^2$-dependence of these transitions.

The CLAS12 detector in Hall B will be a unique facility worldwide,[235] capable of determining transition $\gamma NN^*$ electrocouplings of all prominent excited nucleon states in the almost uncharted region of $Q^2$ from 5-12 GeV$^2$, where $N^*$ structure is expected to be dominated by dressed-quark degrees-of-freedom.[88, 91] CLAS12 will also afford access to parton distributions in an excited nucleon, and enable the concept of GPDs to be applied to the transition of a nucleon to its excited state. New high precision hadro-, photo-, and electroproduction data off the proton and the neutron will stabilize coupled channel analyses and expand the validity of reaction models, enable searches for baryon hybrids and investigation of their structure, establish a repertoire of high precision spectroscopy parameters, and measure light-quark flavor-separated electrocouplings over an extended $Q^2$-range for a wide variety of N* states. Including this body of results in the combined analyses will greatly expand both our knowledge and our understanding of the baryon sector, and the transition from the non-perturbative to the perturbative regime of QCD.

**Parity-violating electron scattering and hadron structure** – Parity-violating (PV) electron scattering and precision neutrino scattering provide additional important information on nucleon and nuclear structure, such as the contribution of strange quarks to nucleon structure and the neutron radius in nuclei, and also enable hadron physics to place constraints on extensions to the Standard Model. Highlights from this program were described in Sec. 2.

The JLab 12 facility promises to deliver new advances, including an ultra-precise measurement of the weak mixing angle, $sin^2\vartheta_W$, via the MOLLER project,[236] which although a low-$Q^2$ experiment will match the precision of the best available collider measurements at the *Z*-boson pole. The SoLID project will employ PVDIS in order to expand sensitivity to beyond the Standard Model weak interaction couplings to a level that will match that of high luminosity experiments at LHC in channels with complementary chiral and flavor combinations.[237]

PVDIS with SoLID will also provide direct sensitivity to parton-level charge-symmetry violation (CSV), which has important implications for PDF fits and could also be part of an explanation for the NuTeV anomaly.[13] With SoLID, PVDIS can deliver a measurement of $d_v/u_v$ at large-*x* that is free from nuclear corrections, which could be critical given the power that a precise measurement of this and similar ratios have for discriminating between competing descriptions of nucleon structure.[97]

**Expressions of chiral dynamics in hadrons and nuclei** – The chiral symmetry of massless QCD is broken dynamically by quark-gluon interactions and explicitly by inclusion of light-quark masses. Consequently, pions and kaons have a special status in QCD and have a marked impact on the long-distance structure of hadrons. This understanding is systematically encoded in χEFT, which is applicable to processes at energies below the chiral symmetry breaking scale via an expansion in a small, dimensionless parameter, thereby allowing an estimation of residual theory uncertainties.

JLab-12 will explore aspects of this physics via measurements of the *γγ* decays of light pseudoscalar mesons ($π^0$, *η*, *η′*) and *η*-*η′* mixing.[238] These processes provide access to the *Abelian* and *non-Abelian chiral anomalies*, which are intimately tied to the pattern of chiral-symmetry breaking. The pion's polarizability also reflects chiral-symmetry breaking in key ways, and it will be measured in Hall-D at JLab-12 via the process[239] $γγ→π^+π^-$. Such measurements assist efforts to calculate the standard-model contribution to the muon's anomalous magnetic moment[240, 241] and measurements of light pseudoscalar meson decays could greatly improve knowledge of the light-quark current-mass ratio $(m_u-m_d)/m_s$.

In connection with hadron polarizabilities, a new generation of experiments with unpolarized and polarized targets (proton, deuteron and $^3$He) and photon beams is approved or planned at HIGS and Mainz. Within the next few years they will provide high-accuracy data on the proton magnetic polarizability,[242, 243] yield precision extractions of spin polarizabilities, and address proton-neutron differences. The monochromatic intensity-frontier laser at HIGS and the tagged bremsstrahlung beam at Mainz complement each other's strengths, and both provide an excellent complement to physics programs with lepton beams. Data extracted from Compton scattering on protons and light nuclei[244] will serve as a benchmark using which, e.g., emerging lattice-QCD calculations[245] of polarizabilities can be validated.

The Gerasimov-Drell-Hearn (GDH) sum rules relate the excitation spectrum of a target's total helicity-dependent photo-absorption cross-section difference to its anomalous magnetic moment. The nucleon sum rules have been tested extensively; but the first studies involving the deuteron and ³He have only recently been carried out.[246, 247, 248] Future experiments at HIGS will test the deuteron sum rule. Progress has also been made in near-threshold $\pi^0$ photo-production, which provides insight into chiral- and isospin symmetry breaking. Single (beam)[249] and double-spin asymmetries in $\pi^0$ photo-production from the proton have been measured and further measurements are planned at both Mainz and HIGS. In the longer term, it is envisioned that parity-violation photodisintegration measurements involving light nuclei will be possible at HIGS.[250] Such measurements are sensitive to the weak force between quarks, whose implications for nuclei have been studied for decades but remain poorly understood.

### 3d. Hadron spectroscopy

Hadron spectroscopy studies the bound states of QCD. The well-known manifestations of these are the baryons (three-quark states) and the mesons (quark-antiquark states). However, from what we know, other color-neutral combinations could also be possible. In terms of quark states, pentaquarks, with the constituents of a baryon and a meson, and four-quark states, with the constituents of a pair of mesons, could be possible. By including valence gluon degrees-of-freedom in addition to the quarks, new objects such as glueballs (glue-only states) and hybrids (a quark object in which the gluonic fields have been excited) are also expected.

**Analysis and Theory** – Advances in hadron spectroscopy require not only new data, but also concerted efforts to more rigorously couple theoretical constraints and expectations into the sophisticated amplitude analyses that are needed to extract new information. With a very strong anchor at Jefferson Lab, over the last 10 years, these efforts have led to significant progress in this area. As remarked in Sec.3a. Hadron theory, there are now close collaborations looking at the best way to unambiguously extract information from new data. These have started to be used in some analyses, and will play an increasingly important role as we move forward into the 12-GeV era with data of unprecedented precision and statistics.

**Baryon Spectroscopy** – Over the last ten years, an intense program of meson photo- and electro-production measurements has resulted in an extensive set of observables. These include not only cross-sections, but also polarization and double-polarization observables. This effort is part of a worldwide program that involves high-profile US participation. The analysis of these new data have begun to contribute toward a resolution of a long-standing issue in baryon spectroscopy; namely, the missing resonance problem.[251, 252, 253, 254] Signals for some of the so-called "missing" states have now been identified in a variety of exclusive channels.[234, 255] However, as illustrated in Figure 14, more data and improved analyses are required before complete understanding may be claimed. In fact, data from the JLab FROST and HD-Ice targets still await analysis, photoproduction data from the neutron are anticipated, and meson beams are being considered. In combination with the analysis of nuclear effects, the neutron data will reveal the isospin structure of baryonic helicity couplings. Notably, lattice-QCD calculations of the baryon spectrum support the existence of the same number of states as expected from constituent-quark models with SU(6) spin-flavor symmetry, so that simple spectroscopic models in which a pair of the quarks are compressed into a pointlike diquark correlation are excluded.[251, 256] This is consistent with Faddeev equation analyses of baryons, which predict instead a measurable role for strong, dynamical and non-pointlike quark-quark correlations.

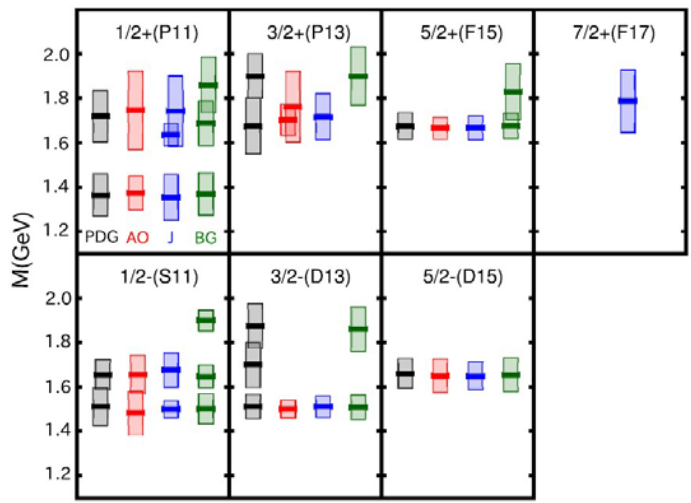

Figure 14 – Three- and four-star nucleon resonance masses as listed by the Particle Data Group[133] and as extracted in three separate analyses: Argonne-Osaka,[253] Jülich[254] and Bonn-Gatchina.[252] For each resonance, Re($M_R$) together with the Re($M_R$) ± Im($M_R$) band is plotted. The four values only agree well in the low-mass region. At higher masses, the differences are large, an outcome that can mainly be attributed to the fact that the available $\pi N$ and $\gamma N$ data for $W \geq 1.7$-GeV reactions are insufficient to determine the partial-wave-amplitudes model independently. Naturally, differences in the analysis methods and the data included in each analysis could also lead to disagreements.

The study of baryons containing two and three strange quarks (the $\Xi$ and $\Omega$ states, respectively) has begun.[257] To date, rather little is known about these states; but with the advent of higher-energy 12-GeV beams, and experiments with complete coverage for charged and neutral particles,[258, 259] we will be able to add significant information to these sectors. The identification and classification of multiply strange baryons will significantly add to our understanding of the manner by which QCD manifests itself in the three-quark arena.

**Intrinsic heavy quarks** – An exciting and important aspect of hadron structure that can be addressed by an EIC are the "intrinsic" strange, charm, and bottom distributions in the proton.[260, 261, 262] In such distributions, the heavy quarks are multi-connected to the proton's valence quarks in the hadron wave function and are thus maximal at equal rapidity and high Bjorken-$x$. Unlike the low x distributions generated by gluon splitting, the intrinsic contributions are charge asymmetric and have strong spin correlations. This remarkable phenomenon can explain many aspects of heavy quark hadroproduction at high $x_F$ and leptoproduction at high Bjorken-$x$. Collider processes such as $pp \to \gamma cX$ at high $p_T$ are critically sensitive to the charm distributions at high $x$.

**Meson Spectroscopy** – In the light-quark meson sector, the emphasis has been on the search for hybrid mesons, where the excited gluonic field contributes directly to the quantum numbers of the meson. A significant result in this area is the lattice QCD calculation of the entire light-quark spectrum, highlighted in Sec. 2, which shows not just the normal $q\bar{q}$ states, but also a number of states with exotic (non $q\bar{q}$) quantum numbers. The lattice simulations indicate that gluonic excitations require an additional 1-GeV of energy. They expose states with exotic quantum numbers, as well as several nonets with normal $q\bar{q}$ quantum numbers.[263] The lightest hybrids with unconventional quantum numbers are confirmed to be those with $J^{PC}=1^{-+}$, well within the GlueX search arena that we discuss below.

Experimentally, most of the new activity has been in the charmonium ($c\bar{c}$) area. A number of narrow $c\bar{c}$ states have been discovered with masses above the $D\bar{D}$ threshold,[264] which cannot be accommodated in the traditional $c\bar{c}$ picture of charmonium states. While none of these has been identified with exotic quantum numbers, it could be that some of them are associated with hybrid mesons. There have also been observations of several charged states that decay to a $c\bar{c}$ state and a charged pion. Of these, one state has been shown to have phase motion consistent with a resonance. In addition, multiple analogous charged states have also been discovered in the bottomonium region, suggesting the same physics is appearing at different quark mass scales. The only way to build these states is to incorporate two quarks and two antiquarks – as a four-quark state, a diquark-antidiquark resonance, or a molecular state. This activity will certainly continue, e.g. at BESIII, Belle-II, and PANDA; and it raises the interesting question of how, if at all, these phenomena map onto the lighter strangeonium ($s\bar{s}$) spectrum.

In the experimental light-quark sector there have been several results on exotic hybrids but most of the activity has been focused on building the next generation of experiments as part of the Jefferson Lab 12-GeV upgrade. Some of the experiments are now poised to take data. Searches for scalar glueball candidates were conducted using central production of two pseudoscalar-meson final states at CERN over a decade ago. Extensions of these studies are underway;[265] and similar measurements might also be possible at RHIC in proton-proton collisions.

**QCD Exotics and Confinement** – As we move into the 12-GeV era at JLab, a major thrust will be the search for light-quark hybrid mesons produced via photo-production. This work will employ the GlueX detector in the new Hall-D as well as the CLAS12 experiment in Hall-B. By measuring exclusive reactions with both charged particles and photons, the statistics expected in this new realm will allow a thorough exploration of meson systems up to masses of about 2.5-GeV, which is a range well matched to expectations based on both models and lattice QCD. A detailed mapping of the experimental spectrum of gluonic excitations, coupled with new phenomenological efforts, and continuum- and lattice-QCD will allow us to understand the role of gluonic fields in the bound states of QCD, which directly couples to the role of these fields in the confinement of quarks. This energy regime also matches what would be expected for strangeonium counterparts to the new charmonium states. With planned kaon-identification upgrades, the search for these states will also be an exciting and important part of the program.

**Understanding bound-states** – With the start of the 12-GeV era at JLab, we will begin an exciting new physics program that has been planned for over a decade. Both theoretical work and measurements carried out over the last 5 years now provide even better guidance concerning where to focus this new program, as well as suggesting interesting opportunities for its extension. At its conclusion, we will understand the evolution of the bound states of QCD from the light-quark into the heavy-quark regime.

### 3e. QCD and nuclei

One of the most important and challenging goals in physics is to understand the fundamental structure of nuclei, the nuclear force and nuclear phenomena from first principles in QCD. Nuclear scientists learned more than two decades ago that the nuclear environment modifies the behavior of quarks and gluons compared to their properties inside an isolated proton or neutron. A notable example is the EMC effect, discovered by the European muon collaboration in late 1980s;[5] namely, the per-nucleon structure functions in nuclei were found to be different from those of the proton determined from lepton deep-inelastic scattering experiments. As highlighted in Sec. 2, precision data and global analyses in the last decade or so have yielded new insights regarding the EMC effect, such as: its potential connection with short-range correlations inside the nucleus; nuclear modification of the quark distributions appearing to depend more on the local nuclear environment; and flavor dependence of nuclear parton distribution functions (PDFs).[13] However, the origin of the EMC effect remains unknown.

The Jefferson Lab 12 GeV Upgrade will both study the QCD structure of nuclei and use the nucleus as a laboratory to study QCD. It will investigate a number of the most fundamental questions in modern nuclear physics:[3]

- What is the nature of the nucleon-nucleon (NN) relative wave function at short distances? Can this be described in terms of nucleons and mesons, or are quarks and gluons necessary?
- The nuclear environment is known to modify the quark-gluon structure of bound nucleons. What is the nature of this modification and how is it related to the short- distance NN wave function?
- How thick is the neutron skin in heavy nuclei? What are the implications for neutron stars?

The nuclear force that binds the proton and neutron into a deuteron (the simplest nucleus) becomes repulsive at short distances, when the proton and neutron are very close to each other. This short-range repulsive behavior is a basic component of all nuclei, required to prevent catastrophic collapse. Thus, the delicate interplay between attraction and repulsion that enables the existence of atomic nuclei, and therefore chemical elements, is a vital topic for current research. A 12-GeV D($e,e'p$)[266] experiment will probe the missing-momentum range 0.5-1.0 GeV, an unprecedented reach, thereby helping to elucidate the nature of both short-range behavior in the nucleon-nucleon force and high density fluctuations in nuclei. The deuteron also remains an interesting laboratory to study non-nucleonic degrees of freedom; e.g., a new experiment[267] employing a tensor polarized deuterium target may reveal non-trivial tensor polarization of the sea through a determination of tensor-polarized deuteron structure functions.

The 12-GeV experimental program will further explore the origin of the EMC effect via a series of SRC and EMC experiments[99, 268, 269, 270] involving different nuclear targets, including those needed to investigate the polarized EMC effect.[12] Inclusive A($e,e'$) scattering, a valuable tool to study nuclei, will be used further in 12-GeV measurements[271] to probe high-momentum nucleons and SRCs in the nucleus at Bjorken-$x$>1, including three-nucleon and four-nucleon SRCs. The highest $Q^2$ data from $x_B > 1$ will probe the distributions of super-fast quarks in nuclei, greatly extending our understanding of nucleons at short distance. It was argued[13] that the interaction of quarks with an isovector-vector mean-field within the nucleus induces differences between *u*- and *d*-quark medium modifications. This flavor dependence of nuclear PDFs can be tested in the SoLID PVDIS measurement[272] from a non-isoscalar target, such as $^{48}$Ca or $^{208}$Pb. It was also argued[38, 273] that quark interactions with scalar mean-fields in nuclei reduce the dynamical quark mass, leading to enhanced lower components of quark spinors. If so, then the spin carried by quarks decreases in nuclei, and the predicted spin EMC effect is even larger than the unpolarized EMC effect. The predicted spin EMC effect will be tested in a new 12-GeV experiment at JLab.[274] These new experiments and theoretical efforts in the coming decade will materially improve our knowledge of the EMC effect and perhaps reveal its origin.

The JLab 12-GeV upgrade will also provide exciting opportunities for studying fundamental processes in the nuclear environment, such as quark propagation and hadronization in the nuclear medium,[275] the possibility of hidden-color degrees of freedom in nuclei,[267, 276, 277, 278] and a search for the onset of the color transparency effect[276, 279] in exclusive processes.

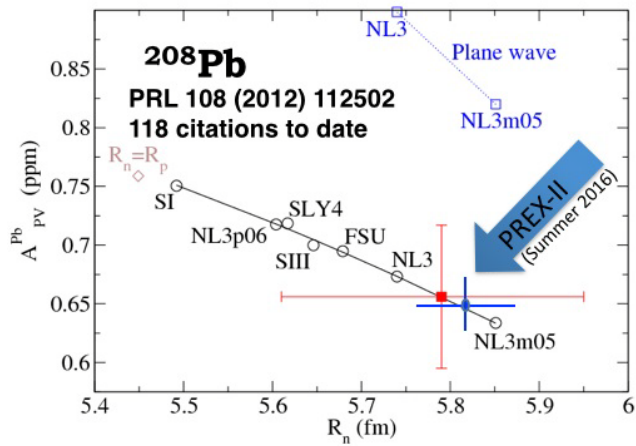

Figure 15 – Projected result from Experiment E12-12-004 (PREX-II) (blue solid circle) together with published result from PREX[280] (red solid square) versus neutron point radius, $R_n$, in $^{208}$Pb. Distorted wave calculations for seven mean-field neutron densities are shown as circles, while the diamonds mark the expectation for $R_n=R_p$. The blue squares show plane wave impulse approximation results.

Parity-violating electron elastic scattering from lead at 6 GeV demonstrated that one can determine, in a model-independent way, the neutron charge density, and provided the first electroweak observation of the neutron skin in a neutron-rich nucleus.[280] As illustrated in Figure 15, a new experiment,[281] on $^{208}$Pb and $^{48}$Ca, following the 12-GeV upgrade will significantly improve the precision of the 6-GeV measurement on $^{208}$Pb and achieve a comparable measurement on $^{48}$Ca. These anticipated results will have significant implications for our understanding of neutron stars.

## 4. Understanding the glue that binds us all: The Electron Ion Collider

### 4a. The Next QCD frontier

Atomic nuclei are built from protons and neutrons, which themselves are composed of quarks that are bound together by gluons. QCD not only determines the structure of hadrons but also provides the fundamental framework to understand the properties and structure of atomic nuclei at all energy scales in the universe. QCD is based on the exchange of gauge bosons, called *gluons*, between the constituents of hadrons, *quarks*. Without gluons there would be no protons, no neutrons, and no atomic nuclei.

Matter as we know it would not exist. Understanding the interior structure and interactions of nucleons and nuclei in terms of the properties and dynamics of the quarks and gluons as dictated by QCD is thus a fundamental and central goal of modern nuclear physics.

Gluons do not carry an electric charge and are thus not *directly* visible to electrons, photons, and other common probes of the structure of matter. An understanding of their role in forming the visible matter in the universe thus remains elusive. The Electron Ion Collider (EIC) with its unique capability to collide polarized electrons with polarized protons and light ions at unprecedented luminosity, and with heavy nuclei at high energy, will be the first precision microscope able to explore how gluons bind quarks to form protons and heavier atomic nuclei.

## 4b. Science highlights and deliverables at the EIC

The high energy, high luminosity polarized EIC will unite and extend the scientific programs at CEBAF and RHIC in dramatic and fundamentally important ways.

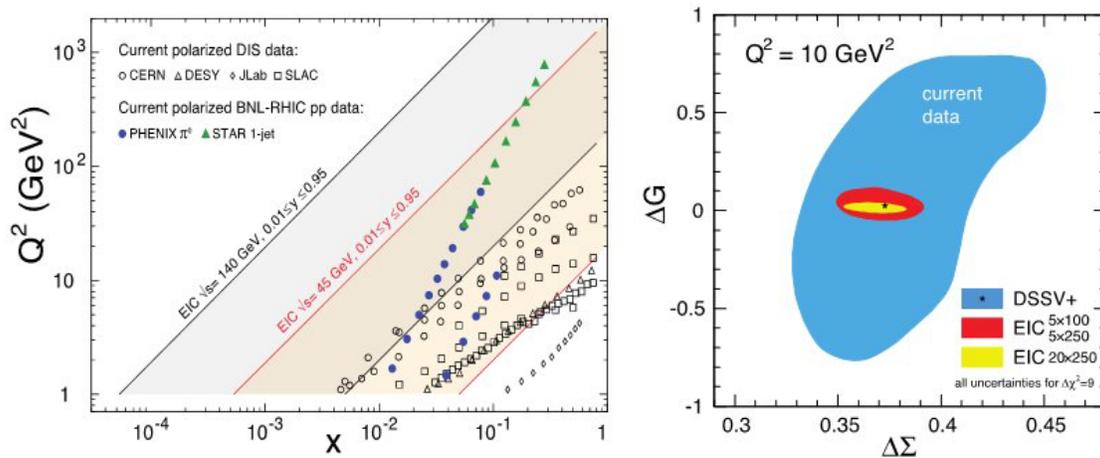

Figure 16 – **Left panel**: The increase in the proton momentum fraction *x* vs. square of the momentum transferred by the electron to proton, $Q^2$, accessible to the EIC in *e+p* collisions. **Right panel**: The projected reduction in the uncertainties of the gluon's ($\Delta G$) and quark's ($\Delta S$) contributions to the proton's spin. The blue band reflects the uncertainties estimated prior to the availability of 2009 data from RHIC. The inclusion of new data published since 2009 does not materially change the overall picture: uncertainties are mainly limited by the lack of low-x data.[282]

**Proton Spin** – Recent measurements at RHIC along with state-of-the-art perturbative QCD analyses have shown that gluons carry approximately 20-30% of the proton's helicity, similar to the quark and anti-quark's contribution. The blue band in the right panel of Figure 16 shows the current level of uncertainties. The knowledge is limited by the *x*-range explored so far. The EIC would greatly increase the kinematic coverage in *x* and $Q^2$, as shown in the left panel of Figure 16, and hence reduce this uncertainty very dramatically, to the level depicted by the red and yellow bands in the right panel.

**Motion of quark and gluons in a proton** – Semi-inclusive measurements with polarized proton beams would enable us to selectively and precisely probe the correlations between the spin of a fast moving proton and the confined transverse motion of both the quarks and gluons within it. Images in momentum space as shown in the left panel of Figure 17 are simply unattainable without the polarized electron and proton beams of the proposed EIC.

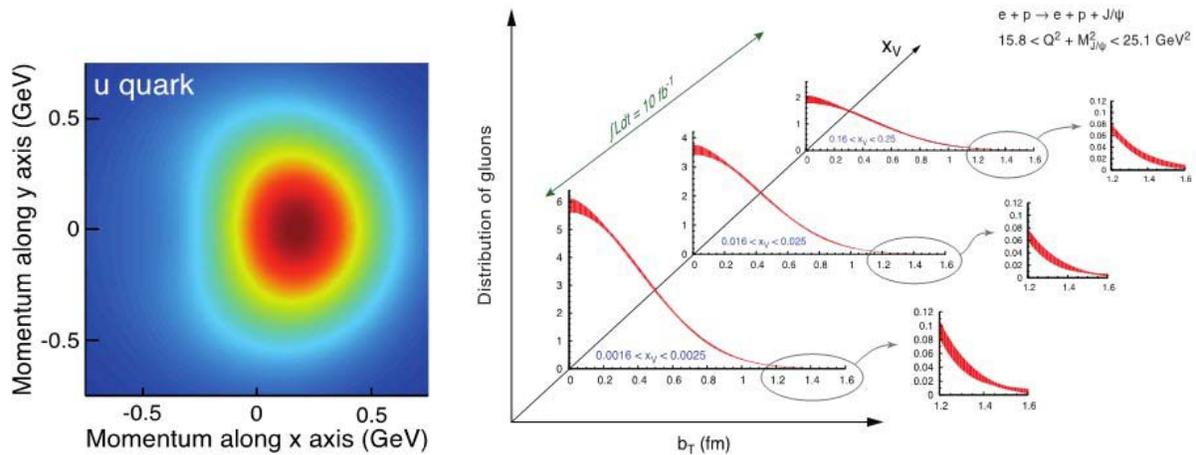

Figure 17 – **Left panel**: Transverse momentum distribution of a *u*- quark with longitudinal momentum fraction *x=0.1* in a transversely polarized proton moving in the *z*-direction, while polarized in *y*-direction. The color code indicates the probability of finding the *u*-quarks, with darker meaning greater probability. **Right panel**: Projected precision of transverse spatial distribution of gluons obtained from exclusive *J/Ψ* production at the EIC.

**Tomographic images of the proton** – By choosing particular final states in $e^-+p$ scattering, the EIC, with its unprecedented luminosity and detector coverage, will create detailed images of the proton's gluon matter distribution, as shown in the right panel of Figure 17. Such measurements would reveal aspects of proton structure that are intimately connected with QCD dynamics at large distances.

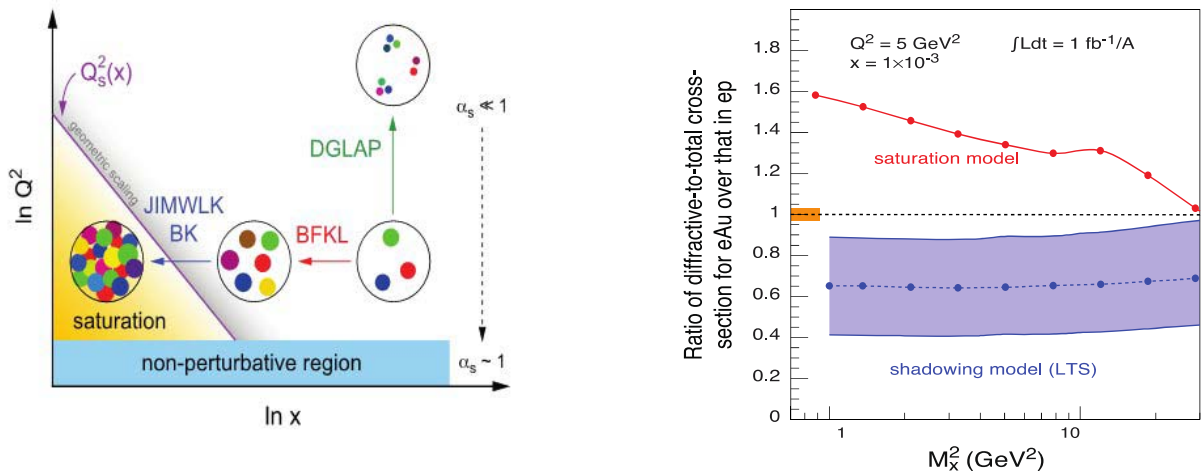

Figure 18 – **Left panel**: Schematic probe resolution vs. energy landscape, indicating regions of non-perturbative and perturbative QCD, including in the latter, low to high parton density and the transition region. **Right panel**: Ratio of diffractive over total cross section for DIS on gold, normalized to DIS on the proton for different value of the mass-squared of hadrons produced in collisions, with and without saturation.

**QCD matter at extreme gluon density** – When fast moving hadrons are probed at high energy, the low-momentum gluons contained in their wave functions become experimentally accessible. By colliding electrons with heavy nuclei moving at light-speed, the EIC will provide access to a so far unconfirmed regime of matter, where abundant gluons dominate its behavior as shown in the left panel of Figure 18. Such universal cold gluon matter is an emergent phenomenon of QCD dynamics and of high scientific interest and curiosity. Furthermore, its properties and its underlying QCD dynamics are critically important for understanding the dynamical origin of the creation of the QGP from colliding two relativistic heavy ions, and the QGP's almost perfect liquid behavior. By measuring diffractive cross-sections together with the total DIS cross-sections in $e^-+p$ and $e^-+A$ collisions, shown in the right panel of Figure 18, the EIC would provide the first unambiguous evidence for the novel QCD matter of saturated gluons. The planned EIC is capable of exploring with precision the new field of collective dynamics of saturated gluons at high energies.

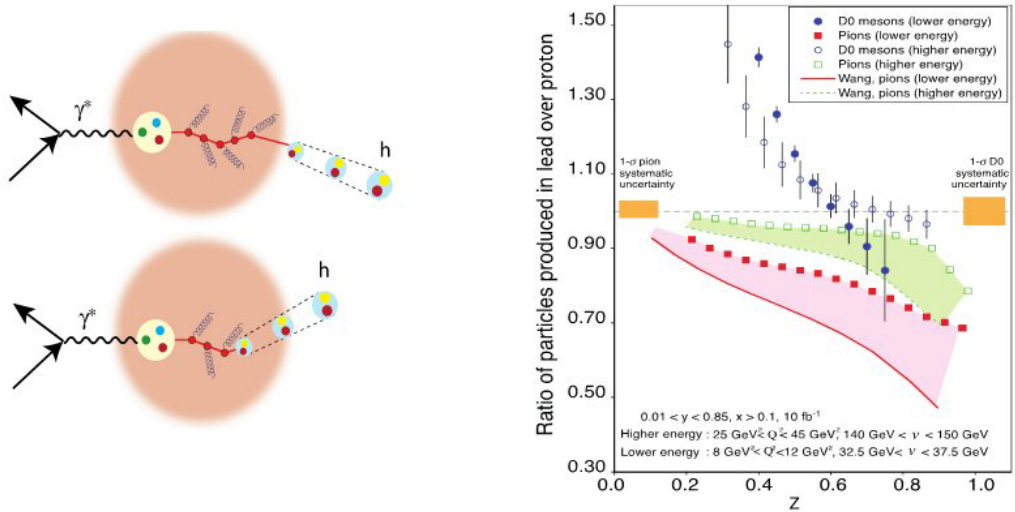

Figure 19 – **Left panel:** A schematic illustrating the interaction of a parton moving through cold nuclear matter, top: hadron is formed outside (top), or inside (bottom) the nucleus. **Right panel:** The ratio of semi-inclusive cross section for producing a pion (light quarks, red) and $D^0$ mesons (heavy quarks, blue) in e+Pb collisions to e+D collisions plotted as a function of z, the ratio of momentum carried by the produced hadron to that of a virtual photon.

**Hadronization and energy loss** – The mechanism by which colored partons pass through colored media, both cold nuclei and hot matter (the QGP), and color-singlet hadrons finally emerge from the colored partons is not understood. A nucleus at the EIC would provide an invaluable femtometer filter with which to explore and expose how colored partons interact and hadronize in nuclear matter, as illustrated in the left panel of Figure 19. By measuring $\pi$ and $D^0$ meson production in both $e^-+p$ and $e^-+A$ collisions, the EIC would provide the first measurement of the quark mass dependence of the response of nuclear matter to a fast moving quark. The dramatic difference between them, shown in the right panel of Figure 19, would readily be discernible. The color bands reflect the limitations on our current knowledge of hadronization – the emergence of a pion from a colored quark. Enabling all such studies in one place, the EIC will be a true "QCD Laboratory", a unique facility in the world.

## 4c. EIC Machine parameters and designs

Two independent designs for the future EIC have evolved over the past few years. Both use existing infrastructure and facilities available to the US nuclear scientists. At Brookhaven National Laboratory (BNL), the eRHIC utilizes a new electron beam facility based on an Energy Recovery Linac (ERL) to be built inside the RHIC tunnel in order to collide electrons with one of the RHIC beams. At Jefferson Laboratory the Medium Energy Electron Ion Collider (MEIC) employs a new electron and ion collider ring complex, together with the 12 GeV upgraded CEBAF in order to achieve similar collision parameters. The machine designs are aimed to achieve:

- Polarized ($\sim$ 70%) beams of electrons, protons and light nuclei;
- Ion beams from deuteron to the heaviest nuclei (uranium or lead);
- Variable center of mass energies from $\sim$ 20-100 GeV, upgradable to $\sim$ 140 GeV;
- High collision luminosity $\sim 10^{33-34}$ cm$^{-2}$sec$^{-1}$;
- Capacity to have more than one interaction region.

### 4d. Why now?

Today, a set of compelling physics questions related to the role of gluons in QCD has been formulated, and a corresponding set of measurements at the EIC identified. A powerful formalism that connects those measurements to the QCD structure of hadrons and nuclei has been developed. The EIC was designated in the 2007 Nuclear Physics Long Range Plan as "*embodying the vision for reaching the next QCD frontier*". In 2013 the NSAC Subcommittee report on Future Scientific Facilities declared an EIC to be "*absolutely essential in its ability to contribute to the world-leading science in the next decade*". Accelerator technology has recently developed so that an EIC with the versatile range of kinematics, beam species and polarization, crucial to addressing the most central questions in QCD, can now be constructed at an affordable cost. Realizing the EIC will be essential to maintain U.S. leadership in the important fields of nuclear and accelerator science.

# Appendix: Agenda of the QCD and Hadron Physics Town Meeting

Here we include an overview of the program. Full details and copies of the presentations are available at https://phys.cst.temple.edu/qcd/.

**Saturday 13 September 2014**

| 08:30-09:00 | ONE – Opening | Chairs: Haiyan Gao and Craig Roberts |
|---|---|---|
| | *Presentation* | *Speaker(s)* |
| 08:30 | Opening remarks: Local Organizing Committee | Prof. Nikos Sparveris (Temple University) |
| 08:45 | Opening remarks: Program Committee | Prof. Haiyan Gao (Duke University), Dr. Craig Roberts (ANL) |

| 09:00-10:30 | TWO: Hadron Structure at Short Distance I | Chair: Sebastian Kuhn |
|---|---|---|
| | *Presentation* | *Speaker(s)* |
| 09:00 | Theoretical Overview: Nucleon spin structure and orbital angular momentum | Dr. Jianwei Qiu (BNL) |
| 09:30 | Experimental overview: Nucleon spin structure | Dr. Ralf Seidl (RIKEN) |
| 10:00 | Parton distribution functions | Prof. Jen-Chieh Peng (University of Illinois at Urbana-Champaign) |

10:30-11:00 – Coffee Break

| 11:00-12:15 | THREE: Hadron Structure at Short Distance II | Chair: Feng Yuan |
|---|---|---|
| | *Presentation* | *Speaker* |
| 11:00 | Nucleon tomography | Prof. Andreas Metz (Temple U.) |
| 11:30 | Pre-Town Meeting summary | Dr. Alexei Prokudin (JLab), Prof. Leonard Gamberg (PSU), Dr. Zhongbo Kang (LANL) |
| 11:36 | SoLID | Dr. Jian-Ping Chen (JLab) |
| 11:42 | Polarized Drell-Yan at FNAL | Prof. Wolfgang Lorenzen (U. Michigan) |
| 11:48 | Submitted presentations and comments from the Community | Dr. M. Liu (LANL), Dr. H. Avakian (JLab), Dr. E. Long (U. New Hampshire), Prof. P. Souder (Syracuse University), Dr. A. Kim (U. Connecticut), Dr. F.-X. Girod (JLab), Prof. B. Surrow (Temple University) |

12:15-13:15 – Lunch Break

| 13:15-15:30 | FOUR: Hadron Structure at Long Distance | Chair: John Arrington |
|---|---|---|
| | *Presentation* | *Speaker* |
| 13:15 | Electromagnetic form factors of nucleons | Prof. Ron Gilman (Rutgers University) |
| 13:40 | Parity-violating electron scattering and hadron structure | Prof. Krishna Kumar (Stony Brook University) |
| 14:05 | Pion form factors | Prof. Tanja Horn (Catholic University of America) |
| 14:25 | Expressions of chiral dynamics in hadrons and nuclei | Prof. Daniel Phillips (Ohio University) |
| 14:45 | Probing Hadron Structure with Photons | Dr. Calvin Howell (Duke University and TUNL) |
| 15:05 | Submitted presentations and comments from the Community | Prof. S. Pate (New Mexico State U.), Dr. M. Mestayer (JLab), Dr. Z. Zhao (ODU/JLab), Prof. B. Norum (U. Virginia), Prof. H. Griesshammer (George Washington U.), Prof. M. Ahmed (North Carolina Central U./ TUNL), Dr. S. Riordan (U. Massachusetts Amherst) |

15:30-16:00 – Coffee Break

| 16:00-18:35 | FIVE: Joint Session of Hadron Physics and QCD with Phases of QCD | Chair: Jim Napolitano |
|---|---|---|
| | *Presentation* | *Speaker* |
| 16:00 | Introduction | Prof. Zein-Eddine Meziani (Temple University) |
| 16:05 | Welcome Address | Dr. Neil Theobald (President, Temple University) |
| 16:15 | Nuclear Theory since 2007 and for the next decade | Prof. David Kaplan (Institute For Nuclear Theory) |
| 16:45 | Community Discussion | |
| 16:55 | Highlights from QCD and Hadron Physics since 2007 | Dr. Rolf Ent (JLab) |
| 17:15 | Vision for QCD and Hadron Physics | Prof. Naomi Makins (University of Illinois at Urbana-Champaign) |
| 17:35 | Community Discussion | |
| 17:45 | RHIC and LHC Overview: Where are we? Where are we going? | Prof. Bill Zajc (Columbia University) |
| 18:25 | Community Discussion | |

18:35-18:45 – Short Break

| 18:45-20:00 | SIX: Joint Session of Hadron Physics and QCD with Phases of QCD | Chair: Richard Milner |
|---|---|---|
| | *Presentation* | *Speaker* |
| 18:45 | Why we need an EIC: a view from 30,000 feet | Prof. Berndt Mueller (BNL / Duke University) |
| 18:45 | Why we need an EIC: a view from 10,000 meters | Prof. Bob McKeown (JLab) |
| 19:25 | Community Discussion | |

20:00 – Adjournment

## Sunday 14 September 2014

| 08:30-11:00 | SEVEN: Joint Session of Hadron Physics and QCD with Phases of QCD | Chair: Jianwei Qiu |
|---|---|---|
| | *Presentation* | *Speaker(s)* |
| 08:30 | Theoretical issues in Nucleon Structure | Prof. Xiangdong Ji (University of Maryland) |
| 09:00 | Probing Nucleon Structure at an EIC | Prof. Zein-Eddine Meziani (Temple University) |
| 09:30 | Probing the properties of QCD with atomic nuclei: theoretical aspects | Prof. Yuri Kovchegov (Ohio State University) |
| 10:00 | Probing the properties of QCD with atomic nuclei: experimental aspects | Dr. Thomas Ullrich (BNL) |
| 11:00 | Community Discussion | |

11:00-11:20 – Coffee Break

| 11:20-12:40 | EIGHT: Joint Session of Hadron Physics and QCD with Phases of QCD | Chair: Craig Roberts |
|---|---|---|
| | *Presentation* | *Speaker(s)* |
| 11:20 | Next-generation nuclear DIS with spectator tagging at EIC | Dr. Christian Weiss (JLab) |
| 11:25 | Neutron Spin Structure via Spectator Tagging at the EIC | Prof. Charles Hyde (Old Dominion University) |
| 11:30 | Transverse momentum dependence of sea quark distributions | Dr. Harut Avakian (JLab) |
| 11:35 | Why QGP Physicists should want to study e+A | Prof. Barbara Jacak (Stony Brook University) |
| 11:40 | EIC White paper discussion | |

12:40-13:40 – Lunch Break

| 13:40-14:45 | NINE: Joint Session of Hadron Physics and QCD with Phases of QCD | Chair: Abhay Deshpande |
|---|---|---|
| | *Presentation* | *Speaker(s)* |
| 13:40 | EIC White paper discussion – and writing assignments | Prof. Abhay Deshpande (SUNY-Stony Brook) |

| 13:40-16:00 | TEN: Joint Session of Hadron Physics and QCD with Phases of QCD | Chair: Paul Sorenson |
|---|---|---|
| | *Presentation* | *Speaker(s)* |
| 14:45 | Report from Computational Nuclear Physics Town Meeting | Dr. Peter Petreczky (BNL) |
| 15:00 | Report from Computational Nuclear Physics Town Meeting | Prof. Martin Savage (Institute for Nuclear Theory) |
| 15:15 | Report from Education and Innovation Town Meeting | Dr. Thia Keppel (JLab) |
| 15:30 | Community Discussion | |

16:00-16:30 – Coffee Break

| 16:30-18:30 | ELEVEN: QCD and Hadron Physics/Theory | Chair: Craig Roberts |
|---|---|---|
| | *Presentation* | *Speaker(s)* |
| 16:30 | Lattice-based studies of QCD | Dr. D.G. Richards (JLab) |
| 17:00 | Continuum-based studies of QCD | Dr. Ian Cloët (ANL) |
| 17:30 | pQCD at the collider (LHC/RHIC) | Prof. Iain Stewart (MIT) |
| 18:00 | Submitted presentations and comments from the Community | Dr. C. Lee (LANL), Prof. S. Liuti (U. Virginia), Prof. B. Tiburzi (City College of New York), Prof. M. Burkardt (NMSU) |

18:30 – Adjournment

## Monday 15 September 2014

| 08:30-10:00 | TWELVE: Hadron Spectroscopy I | Chair: Volker Burkert |
|---|---|---|
| | *Presentation* | *Speaker(s)* |
| 08:30 | Theory Overview: QCD and Meson spectrum | Dr. Michael Pennington (JLab) |
| 09:00 | Experiment Overview on Meson Spectroscopy: present and future | Prof. Curtis Meyer (Carnegie Mellon University) |
| 09:30 | Theory Overview on Baryon Spectroscopy | Dr. Michael Döring (George Washington University) |

10:00-10:20 – Coffee Break

| 10:20-11:10 | THIRTEEN: Hadron Spectroscopy II | Chair: Patrizia Rossi |
|---|---|---|
| | *Presentation* | *Speaker(s)* |
| 10:20 | Experimental overview of Baryon Spectroscopy: present and future | Prof. Ralf Gothe (University of South Carolina) |
| 10:45 | Submitted presentations and comments from the Community | Prof. W. Briscoe (GWU), Prof. Igor Strakovsky (GWU), Prof. Lei Guo (Florida International U.), Dr. Victor Mokeev (JLab), Dr. Ryan Mitchell (Indiana U.), Prof. Adam Szczepaniak (Indiana U. and JLab) |

| 11:10-13:00 | FOURTEEN: QCD and Nuclei | Chair: Haiyan Gao |
|---|---|---|
| | *Presentation* | *Speaker(s)* |
| 11:10 | Effect of short-range interactions on nuclei | Dr. Doug Higinbotham (JLab) |
| 11:40 | Nuclei at short distance scales | Prof. Patricia Solvignon (University Of New Hampshire) |
| 12:05 | Parton distributions in nuclei | Dr. Wally Melnitchouk (JLab) |
| 12:35 | Submitted presentations and comments from the Community | Prof. Henry Weller (Duke U. and TUNL) |

13:00-14:00 – Lunch Break

| 14:00-16:00 | FIFTEEN: Closing Session | Chairs: Haiyan Gao and Craig Roberts |
|---|---|---|
| | *Presentation* | *Speaker(s)* |
| 14:00 | Presentation and discussion of Recommendations | Haiyan Gao and Craig Roberts |

16:00 – End of Town Meeting

## Acknowledgments

Comments and suggestions from many scientists in the Hadron Physics community were extremely valuable in the course of editing this summary, and are gratefully acknowledged.

# References

[1] The Frontiers of Nuclear Science: A long range plan, R. Tribble *et al*., (NSAC Web Link to LRP)
[2] Report to NSAC on Implementing the 2007 Long Range Plan (NSAC_FacilitiesReport.pdf)
[3] J. Dudek *et al*., Eur. Phys. J. A **48** (2012) 187
[4] A. Accardi *et al*., arXiv:1212.1701 [nucl-ex]
[5] European Muon Collaboration (J. J. Aubert *et al.*), Phys. Lett. B 123 (1983) 275
[6] J. Seeley *et al*, Phys. Rev. Lett. **103** (2009) 202301
[7] R. Subedi *et al*., Science **320** (2008) 1476
[8] O. Hen *et al*., Science **346** (2014) 614
[9] L. Weinstein *et al*, Phy. Rev. Lett. **106** (2011) 052301
[10] J. Arrington *et al*., Phys. Rev. C **86** (2012) 065204
[11] N. Fomin *et al*., Phys. Rev. Lett. **108** (2012) 092502
[12] I.C. Cloët *et al.*, Phys. Lett. B **642** (2006) 210
[13] I.C. Cloët *et al.*, Phy. Rev. Lett. **102** (2009) 252301
[14] R. Pohl *et al*., Nature **466** (2010) 213
[15] A. Antognini *et al*., Science **339** (2013) 417
[16] A1 Collaboration (J.C. Bernauer *et al.)*, Phys. Rev. Lett. **105** (2010) 242001
[17] X. Zhan *et al*., Phys. Lett. B **705** (2011) 59
[18] P.J. Mohr, B.N. Taylor, D.B. Newell, Rev. Mod. Phys. **80** (2008) 633
[19] D. Dutta *et al*., JLab Experiment E12-11-106
[20] I. Bernauer and M. Distler, private communication.
[21] MUSE Collaboration (R. Gilman *et al*.), arXiv:1303.2160 [nucl-ex]
[22] EBAC Collaboration (N. Suzuki *et al*.), Phys. Rev. Lett. **104** (2010) 042302
[23] CLAS Collaboration (I.G. Aznauryan *et al*.) Phys. Rev. C **80** (2009) 055203
[24] CLAS Collaboration (V. I. Mokeev *et al*.), Phys. Rev. C **86** (2012) 035203
[25] I.G. Aznauryan and V.D. Burkert, Phys. Rev C **85** (2012) 055202
[26] I.C. Cloët *et al*., Few Body Syst. **46** (2009) pp. 1-36
[27] J. Segovia, I.C. Cloët, C.D. Roberts and S.M. Schmidt, Few Body Syst. **55** (2014) 1185-1222
[28] M. Guidal, M. V. Polyakov, A. V. Radyushkin, and M. Vanderhaeghen, Phys. Rev. D **72** (2005) 054013
[29] I.C. Cloët and G.A. Miller, PRC **86**, 015208 (2012)
[30] I.C. Cloët, W. Bentz and A.W. Thomas, Phys. Rev. C **90** (2014) 045202
[31] S. Riordan *et al*. Phys. Rev. Lett. **105** (2010) 262302
[32] G.D. Cates, C.W. de Jager, S. Riordan and B. Wojtsekhowski, Phys. Rev. Lett. **106** (2011) 252003
[33] I.A. Qattan and J. Arrington, Phys. Rev. C **86** (2012) 065210
[34] S. Gilad *et al.,* Jefferson Lab Experiment E12-07-108
[35] E. Brash *et al.,* Jefferson Lab Experiment E12-07-109
[36] G. Gilfoyle *et al.,* Jefferson Lab Experiment E12-07-104
[37] B. Quinn *et al.,* Jefferson Lab Experiment E12-09-019
[38] G. Cates *et al.,* Jefferson Lab Experiment E12-09-016
[39] B. Anderson *et al.,* Jefferson Lab Experiment E12-11-009
[40] JLab HAPPEX Collaboration (Z. Ahmed *et al*.), Phys. Rev. Lett. 108 (2012) 102001
[41] JLab Qweak Collaboration (D. Androic *et al*.), Phys. Rev. Lett. 111 (2013) 141803
[42] JLab PVDIS Collaboration (D. Wang *et al*.), Nature 506 (2014) 67
[43] *Nuclear Physics: Exploring the Heart of Matter*, The Committee on the Assessment of and Outlook for Nuclear Physics; Board on Physics and Astronomy; Division on Engineering and Physical Sciences; National Research Council (National Academies Press, Washington, D.C. 2013)
[44] L. Chang *et al*., Phys. Rev. Lett. **110** 132001 (2013) [5 pages]
[45] S. V. Mikhailov and A. V. Radyushkin, JETP Lett. **43**, 712 (1986)
[46] V. Y. Petrov, *et al*., Phys. Rev. D **59**, 114018 (1999)
[47] V. M. Braun *et al*., Phys. Rev. D **74**, 074501 (2006)

[48] S. J. Brodsky and G. F. de Teramond, Phys. Rev. Lett. **96**, 201601 (2006)
[49] I. C. Cloët *et al.*, Phys. Rev. Lett. **111** (2013) 092001 [5 pages]
[50] L. Chang *et al.*, Phys. Rev. Lett. **111** (2013) 141802 [5 pages]
[51] G.M. Huber, D. Gaskell *et al.*, Jefferson Lab Experiment E12-06-101
[52] S.R. Amendolia, Nucl. Phys. **B** 277 (1986) 168
[53] G. Huber *et al.*, Phys. Rev. **C** 78 (2008) 045203
[54] T. Horn, G.M. Huber *et al.*, Jefferson Lab Experiment E12-07-105
[55] BaBar Collaboration (B. Aubert *et al.*), Phys. Rev. D **80** (2009) 052002
[56] Belle Collaboration (S. Uehara *et al.*), Phys. Rev. D **86** (2012) 092007
[57] Hadron Spectrum Collaboration (J.J. Dudek *et al.*), Phys. Rev. **D** 88 (2013) 094505
[58] CLAS Collaboration (K. Moriya et al.), Phys. Rev. Lett. **112** (2014) 082004
[59] STAR Collaboration (L. Adamczyk *et al.*), arXiv:1405.5134
[60] PHENIX Collaboration (A. Adare *et al.*), Phys. Rev. D 90 (2014) 012007.
[61] D. de Florian, R. Sassot, M. Stratmann and W. Volgelsang, Phys. Rev. Lett. **113** (2014) 012001
[62] J.-P. Chen, A. Deur, S. Kuhn and Z.-E. Meziani, J. Phys. Conf. Ser. **299** (2011) 012005
[63] M. Posik *et al.*, Phys. Rev. Lett. **113** (2014) 022002
[64] A. Deur *et al.*, Phys. Rev. C **90** (2014) 012009
[65] LHPC Collaboration (Ph. Hägler *et al.*), Phys.Rev. D **77** (2008) 094502
[66] A.W. Thomas, Phys. Rev. Lett. **101** (2008) 102003
[67] LHPC Collaboration (J.D. Bratt *et al.*), Phys.Rev. D **82** (2010) 094502
[68] QCDSF Collaboration (G.S. Bali *et al.*) Phys. Rev. Lett. **108** (2012) 222001
[69] P.E. Shanahan *et al.*, Phys. Rev. Lett. **110** (2013) 20, 202001
[70] A.J. Chambers *et al.* Phys.Rev. D **90** (2014) 014510
[71] M. Deka *et al.*, arXiv:1312.4816 [hep-lat]
[72] X. Ji, Phys. Rev. Lett. **78** (1997) 610-613
[73] A. Airapetian *et al.*, Phys. Rev. Lett. **94** (2005) 012002.
[74] M. Alekseev et al., Phys. Lett. B 673 (2009) 127; Phys. Lett. B 692 (2010) 240
[75] X. Qian *et al*, Phys. Rev. Lett. 107 (2011) 072003; J. Huang *et al.*, Phys. Rev. Lett. 108 (2012) 052001
[76] Belle Collaboration (K. Abe *et al.*), Phys. Rev. Lett. **96** (2006) 232002
[77] Belle Collaboration (R. Seidl *et al.*), Phys. Rev. D **78** (2008) 032011, *Erratum-ibid*. D **86** (2012) 039905
[78] BaBar Collaboration (I. Garzia for the collaboration), Nuovo Cim. C035N2 (2012) 79-84
[79] D. de Florian *et al.*, Phys. Rev. Lett. **101** (2008) 072001; Phys. Rev. D **80** (2009) 034030
[80] A. Walker-Loud, C.E. Carlson and G.A. Miller, Phys. Rev. Lett. **108** (2012) 232301
[81] G.A. Miller, Phys. Lett. B **718** (2013) 1078-1082
[82] J.A. McGovern, D.R. Phillips and H.W. Griesshammer, Eur. Phys. J. A **49** (2013) 12
[83] COMPTON@MAX-lab Collaboration (L.S. Myers *et al.*), arXiv:1409.3705 [nucl-ex]
[84] P.P. Martel et al., arXiv:1408.1576
[85] *The Millenium Prize Problems*, eds. J. Carlson, A. Jaffe and A. Wiles (American Mathematical Society, Providence, 2006)
[86] S.X. Qin *et al.*, Phys. Rev. C **84** (2011) 042202(R) [5 pages]
[87] A.C. Aguilar, D. Binosi and J. Papavassiliou, Phys. Rev. D **88** (2013) 074010
[88] I.C. Cloët and C.D. Roberts, Prog. Part. Nucl. Phys. **77** (2014) 1–69
[89] D. Binosi, L. Chang, J. Papavassiliou and C.D. Roberts, arXiv:1412.4782 [nucl-th]
[90] I.C. Cloët, C.D. Roberts and A.W. Thomas, Phys. Rev. Lett. **111** (2013) 101803 [5 pages]
[91] I.G. Aznauryan *et al.*, Intern. J. Mod. Phys. E **22** (2013) 1330015
[92] C.E. Carlson, Phys. Rev. D **34** (1986) 2704
[93] I.G. Aznauryan and V.D. Burkert, Prog. Part. Nucl. Phys. **67** (2012) 1-54
[94] V. Pascalutsa, M. Vanderhaeghen and S.N. Yang, Phys. Rept. **437** (2007) 125-232
[95] C. Alexandrou, C.N. Papanicolas and M. Vanderhaeghen, Rev. Mod. Phys. **84** (2012) 1231
[96] R.J. Holt and C.D. Roberts, Rev. Mod. Phys. **82** (2010) 2991-3044
[97] C.D. Roberts, R.J. Holt and S.M. Schmidt, Phys. Lett. B **727** (2013) pp. 249–254

[98] BoNuS Collaboration (S. Bueltmann *et al.*), JLab Experiment E12-06-113

[99] MARATHON Collaboration (G. Petratos *et al.*), JLab Experiment E12-10-103

[100] SoLID Collaboration (P. Souder *et al.*), JLab Experiment E12-10-007

[101] CLAS Collaboration (S. Khun *et al.*), JLab Experiment E12-06-109

[102] X. Zheng *et al.*, JLab Experiment E12-06-110

[103] N. Liyanage *et al.*, JLab Experiment E12-06-122

[104] J.J. Dudek *et al.*, Phys. Rev. Lett. **103** (2009) 262001

[105] J.J. Dudek *et al.* Phys. Rev. D **82** (2010) 034508

[106] J.J. Dudek *et al.*, Phys. Rev. D **83** (2011) 111502

[107] LHPC Collaboration (Ph. Hägler *et al.*) Phys. Rev. D **77** (2008) 094502

[108] LHPC Collaboration (J.D. Bratt *et al.*) Phys. Rev. D **82** (2010) 094502

[109] J.R. Green *et al.*, Phys. Lett. B **734** (2014) 290-295

[110] B.U. Musch, Ph. Hägler, J.W. Negele and A. Schäfer, Phys. Rev. D **83** (2011) 094507

[111] B.U. Musch *et al.*, Phys. Rev. D **85** (2012) 094510

[112] S.R. Beane *et al.*, Phys. Rev. Lett. **113** (2014) 252001

[113] S.R. Beane *et al.*, Phys.Rev. D **87** (2013) 3, 034506

[114] Hadron Spectrum Collaboration (J.J. Dudek *et al.*) Phys. Rev. Lett. **113** (2014) 182001

[115] D.J. Wilson, J.J. Dudek, R.G. Edwards and C.E. Thomas, arXiv:1411.2004 [hep-ph]

[116] R.A. Briceño, M.T. Hansen and A. Walker-Loud, arXiv:1406.5965 [hep-lat]

[117] X. Ji, Phys. Rev. Lett. **110** (2013) 262002

[118] J.R. Green *et al.*, Phys. Rev. D **90** (2014) 7, 074507

[119] NPLQCD Collaboration (S.R. Beane *et al.*) Phys. Rev. D **85** (2012) 034505

[120] M. Benayoun *et al.*, arXiv:1407.4021 [hep-ph]

[121] A. Calle-Cordon, T. DeGrand and J.L. Goity, Phys. Rev. D **90** (2014) 014505

[122] C.W. Bauer, S. Fleming and M.E. Luke, Phys. Rev. D **63** (2000) 014006

[123] C.W. Bauer, S. Fleming, D. Pirjol and I.W. Stewart, Phys. Rev. D **63**, 114020 (2001)

[124] C.W. Bauer and I.W. Stewart, Phys. Lett. B **516**, 134 (2001)

[125] C.W. Bauer, D. Pirjol and I. W. Stewart, Phys. Rev. D **65** (2002) 054022

[126] J. Blümlein, H. Bottcher and A. Guffanti, Nucl. Phys. B **774** (2007) 182-207

[127] HPQCD Collaboration (C.T.H. Davies *et al.*), Phys. Rev. D **78** (2008) 114507

[128] H. Flacher et al., Eur. Phys. J. C **60** (2009) 543-583, *Erratum-ibid*. C **71** (2011) 1718

[129] M. Beneke and M. Jamin, JHEP 0809 (2008) 044

[130] M. Davier, S. Descotes-Genon, A. Hocker (CERN), B. Malaescu and Z. Zhang, Eur. Phys. J. C **56** (2008) 305-322

[131] S. Bethke, Prog. Part. Nucl. Phys. **58** (2007) 351-386

[132] S. Bethke, Eur. Phys. J. C **64** (2009) 689-703

[133] Particle Data Group (J. Beringer *et al.*), Phys. Rev. D **86**, 010001 (2012)

[134] S. Fleming, A.K. Leibovich and T. Mehen, Phys. Rev. D **68** (2003) 094011

[135] A. Manohar, Phys. Rev. D **68** (2003) 114019

[136] M. Neubert, Eur. Phys. J. C **40** (2005) 165

[137] T. Becher and M. D. Schwartz, JHEP 0807 (2008) 034

[138] S. Fleming, A.H Hoang, S. Mantry and I.W. Stewart, Phys.Rev. D **77** (2008) 114003

[139] S. Mantry and F. Petriello, Phys. Rev. D **81** (2010) 093007

[140] S.D. Ellis *et al.*, Phys. Lett. B **689** (2010) 82

[141] M.G. Echevarria, Ahmad Idilbi and I. Scimemi, JHEP 1207 (2012) 002

[142] I.W. Stewart, F.J. Tackmann and W.J. Waalewijn, Phys. Rev. Lett. **105** (2010) 092002

[143] Z.B. Kang, S. Mantry and J.W. Qiu, Phys. Rev. D **86** (2012) 114011

[144] S. Fleming, A.K. Leibovich, T. Mehen and I.Z. Rothstein, Phys. Rev. D **86** (2012) 094012

[145] D. Krohn, T. Lin, M.D. Schwartz and W.J. Waalewijn, Phys. Rev. Lett **110** (2013) 212001

[146] D.Kang, C.Lee and I.W. Stewart, Phys.Rev. D **88** (2013) 054004

[147] A. Larkoski, D. Neill and J. Thaler, JHEP 1404 (2014) 017

[148] M. Baumgart, A.K. Leibovich, T. Mehen and I.Z. Rothstein, JHEP 1411 (2014) 003

[149] A. Gehrmann-De Ridder, T. Gehrmann, E.W.N. Glover and G. Heinrich, JHEP 0711 (2007) 058
[150] S. Weinzierl, Phys. Rev. Lett. **101** (2008) 162001
[151] Z. Bern *et al.*, Phys.Rev. D **88** (2013) 1, 014025
[152] BlackHat Collaboration (Z. Bern *et al.*), arXiv:1407.6564 [hep-ph]
[153] A. Gehrmann-De Ridder, T. Gehrmann, E.W.N. Glover and G. Heinrich, Phys. Rev. Lett. **99** (2007) 132002
[154] S. Weinzierl, JHEP 0906 (2009) 041
[155] A. Gehrmann-De Ridder, T. Gehrmann, E.W.N. Glover and J. Pires, Phys. Rev. Lett. **110** (2013) 162003
[156] R. Boughezal, F. Caola, K. Melnikov and F. Petriello, JHEP 1306 (2013) 072
[157] H. Abramowicz *et al.*, arXiv:1306.5413 [hep-ph]
[158] T. Becher and M.D. Schwartz, JHEP 0807 (2008) 034
[159] R. Abbate *et al.*, Phys. Rev. D **83** (2011) 074021
[160] Z.B. Kang, X. Liu and S. Mantry, Phys. Rev. D **90** (2014) 014041
[161] D. Kang, C. Lee and I.W. Stewart, JHEP 1411 (2014) 132
[162] D. Kang, C. Lee and I.W. Stewart, private communication
[163] A. Idilbi and A. Majumder, Phys. Rev. D **80** (2009) 054022
[164] F. D'Eramo, H. Liu and K. Rajagopal, Phys. Rev. D **84** (2011) 065015
[165] G. Ovanesyan and I. Vitev, JHEP 1106 (2011) 080
[166] Y.T. Chien and I. Vitev, arXiv:1405.4293 [hep-ph]
[167] S. Moch, J.A.M. Vermaseren and A. Vogt, Nucl. Phys. B **889** (2014) 351-400
[168] M. Dittmar *et al.* arXiv:0901.2504 [hep-ph]
[169] CMS Collaboration (S. Chatrchyan *et al.*), Phys.Rev. D **87** (2013) 11, 112002
[170] S.M. Aybat, J.C. Collins, J.W. Qiu and T.C. Rogers, Phys. Rev. D **85** (2012) 034043
[171] S.M. Aybat, A. Prokudin and T.C. Rogers, Phys. Rev. Lett. **108** (2012) 242003
[172] M.G. Echevarria, A. Idilbi, A. Schäfer and Ignazio Scimemi, Eur. Phys. J. C **73** (2013) 2636
[173] P. Sun and F. Yuan, Phys. Rev. D **88** (2013), 114012
[174] M.G. Echevarria, A. Idilbi, Z.B. Kang and I. Vitev, Phys. Rev. D **89** (2014) 074013
[175] J.C. Collins and T. Rogers, arXiv:1412.3820 [hep-ph]
[176] J.C. Collins and J.W. Qiu, Phys. Rev. D **75** (2007) 114014
[177] T.C. Rogers and P.J. Mulders, Phys. Rev. D **81** (2010) 094006
[178] R.L. Workman, M.W. Paris, W.J. Briscoe and I.I. Strakovsky, Phys. Rev. C **86** (2012) 015202
[179] M. Shrestha and D.M. Manley, Phys. Rev. C **86** (2012) 055203
[180] S.J. Brodsky, G.F. de Teramond, H.G. Dosch and J. Erlich, arXiv:1407.8131 [hep-ph]
[181] X. d. Ji, Phys. Rev. Lett. **91**, 062001 (2003)
[182] A. V. Belitsky, X. Ji and F. Yuan, Phys. Rev. D **69**, 074014 (2004)
[183] S. Meissner, A. Metz and M. Schlegel, JHEP 0908 (2009) 056
[184] C. Lorcé and B. Pasquini, Phys. Rev. D **84** (2011) 014015
[185] H. Avakian *et al.* Mod. Phys. Lett. A **24** (2009) 2995-3004
[186] M. Guidal, H. Moutarde and M. Vanderhaeghen, Rept. Prog. Phys. **76** (2013) 066202
[187] H. Avakian *et al.*, Jefferson Lab Experiment E12-06-112
[188] H. Avakian *et al.*, Jefferson Lab Experiment E12-07-107
[189] H. Avakian *et al.*, Jefferson Lab Experiment E12-09-008
[190] H. Avakian *et al.*, Jefferson Lab Experiment E12-09-009
[191] R. Ent *et al.*, Jefferson Lab Experiment E12-09-017
[192] G. Cates *et al.*, Jefferson Lab Experiment E12-09-018
[193] J.P. Chen *et al.*, Jefferson Lab Experiment E12-10-006
[194] J.P. Chen *et al.*, Jefferson Lab Experiment E12-11-007
[195] H. Avakian *et al.*, Jefferson Lab Experiment C12-11-111
[196] H. Avakian *et al.*, Jefferson Lab Experiment C12-12-009
[197] S. Mert Aybat and T.C. Rogers, Phys. Rev. D **83** (2011) 114042
[198] P. Sun and F. Yuan, Phys. Rev. D **88** (2013) 034016
[199] M. Anselmino *et al.*, , Phys. Rev. D **87** (2013) 094019

[200] M. Anselmino *et al*., Nucl. Phys. B, Proc. Suppl. **191** (2009) 98
[201] A. Bacchetta, A. Courtoy, M. Radici, JHEP **1303** (2013) 119
[202] A. Prokudin, private communications
[203] ETM Collaboration (C. Alexandrou *et al*.), to appear in PoS(Lattice 2014) **151**
[204] M. Göckeler et al., Phys. Lett. B 627 (2005) 113
[205] M. Pitschmann, C-Y. Seng, C. Roberts, S.M. Schmidt, arXiv:1411.2052
[206] M.B. Hecht, C.D. Roberts, and S. M. Schmidt et al., Phys. Rev. C **64** (2001) 025204
[207] I.C. Clöet, W. Bentz and A.W. Thomas, Phys. Lett. B 659 (2008) 214
[208] B. Pasquini, M. Pincetti and S. Boffi, Phys. Rev. D **76** (2007) 034020
[209] M. Wakamatsu, Phys. Lett. B **653** (2007) 398
[210] L. Gamberg and G.R. Goldstein, Phys. Rev. Lett. **87** (2001) 242001
[211] H. He and X.D. Ji, Phys. Rev. D **52** (1995) 2960
[212] R. Ent *et al*., Jefferson Lab Experiment E12-13-007
[213] T. Horn *et al*., Jefferson Lab Experiment E12-13-010
[214] Z.B. Kang, J.W. Qiu, W. Vogelsang and F. Yuan, Phys. Rev. D **83**, 094001 (2011)
[215] J.C. Collins, Phys. Lett. B **536** (2002) 43-48
[216] S.J. Brodsky, D.S. Hwang and I. Schmidt, Nucl. Phys. B **642** (2002) 344-356
[217] S. J. Brodsky *et al*. Phys. Rev. D **88** (2013) 1, 014032
[218] E.C. Aschenauer *et al*., arXiv:1304.0079 [nucl-ex]
[219] CLAS Collaboration (J. Lachniet *et al*.), Phys. Rev. Lett. **102** (2009) 192001
[220] S. Riordan *et al*. Phys.Rev.Lett. **105** (2010) 262302
[221] A.J.R. Puckett *et al*., Phys. Rev. Lett. **104** (2010) 242301
[222] A.J.R. Puckett *et al*., Phys. Rev. C **85** (2012) 045203
[223] Jefferson Lab Hall A Collaboration (G. Ron *et al*.) Phys. Rev.  C **84** (2011) 055204
[224] A1 Collaboration (J.C. Bernauer *et al*.), Phys. Rev. C **90** (2014) 015206
[225] MicroBooNE Experiment: http://www-microboone.fnal.gov/
[226] S. Pate and D. Trujillo, EPJ Web Conf. **66** (2014) 06018
[227] J.R. Ellis, K.A. Olive and C. Savage, Phys. Rev. D **77** (2008) 065026
[228] BaBar Collaboration (B. Aubert *et al*.) Phys. Rev. D **80** (2009) 052002
[229] Belle Collaboration (S. Uehara *et al*.), Phys.Rev. D **86** (2012) 092007
[230] Jefferson Lab Fpi Collaboration (G.M. Huber et al.), Phys. Rev. Lett. **112** (2014) 182501
[231] S.V. Goloskokov and P. Kroll, Eur. Phys. J. A **47** (2011) 112
[232] T. Horn, Phys. Rev. C **85** (2012) 018202
[233] C. Shi et al., Phys. Lett. B **738** (2014) 512–518
[234] I. Aznauryan, V.D. Burkert, T.S.H. Lee and V.I. Mokeev, J. Phys. Conf. Ser. **299** (2011) 012008
[235] R. Gothe *et al.*, Jefferson Lab Experiment E12-09-003
[236] K.S. Kumar *et al.*, JLab Experiment E12-09-005
[237] J. Erler, C.J. Horowitz, S. Mantry and P.A. Souder, Ann. Rev. Nucl. Part. Sci. **64** (2014) 269-298
[238] A. Gasparian *et al*., Jefferson Lab experiment E12-10-011
[239] R. Miskimen *et al*., Jefferson Lab experiment E12-13-008
[240] K.T. Engel and M.J. Ramsey-Musolf, Phys. Lett. B **738** (2014) 123-127
[241] G. Colangelo *et al*., Phys. Lett. B **738** (2014) 6-12
[242] V. Sokhoyan and E.J. Downie, private communications
[243] M. Ahmed and H.R. Weller, private communications
[244] H.W. Griesshammer, J.A. McGovern, D.R. Phillips and G. Feldman, Prog. Part. Nucl. Phys. **67** (2012) 841-897
[245] M. Lujan, A. Alexandru, W. Freeman and F. Lee, arXiv:1411.0047 [hep-lat]
[246] M.W. Ahmed *et al*., Phys. Rev. C **77** (2008) 0444005; M.A. Blackston *et al*., Phys. Rev. C **78** (2008) 034003
[247] G. Laskaris *et al*., Phys. Rev. Lett. **110** (2013) 202501; Phys. Rev. C **89** (2014) 024002
[248] P. Aguar Bartolomé *et al*., Phys. Lett. B 723 (2013) 71
[249] D. Hornidge *et al*., Phys. Rev. Lett. **111** (2013) 062004
[250] M.W. Ahmed *et al*., arXiv: 1307.8178

[251] V. Crede *et al.*, Rep. Prog. Phys. **76** (2013) 076301
[252] A.V. Anisovich *et al.*, Eur. Phys. J. A **48** (2012) 15, Eur. Phys. J. A **48** (2012) 88
[253] H. Kamano, S.X. Nakamura, T.S. H. Lee and T. Sato , Phys. Rev. C **88** (2013) 035209
[254] D. Rönchen *et al.*, Eur. Phys. J. A **49** (2013) 44
[255] V.D. Burkert, Int. J. Mod. Phys. Conf. Ser. **26** (2014) 1460050
[256] R. Edwards *et al.*, PRD **84**, 074508 (2011), Phys. Rev. D **87** (2013) 054506
[257] CLAS Collaboration (L. Guo *et al.*), Phys. Rev. C **76** (2007) 025288
[258] L. Guo *et al.*, Jefferson Lab experiment E12-11-005A
[259] C.A. Meyer *et al.*, Jefferson Lab experiment E12-13-003
[260] S.J. Brodsky, P. Hoyer, C. Peterson and N. Sakai, Phys. Lett. B **93** (1980) 451-455
[261] S.J. Brodsky, C. Peterson and N. Sakai, Phys. Rev. D **23** (1981) 2745
[262] W.C. Chang and J.C. Peng, Phys. Rev. Lett. **106** (2011) 252002
[263] C.A. Meyer *et al.*, Phys. Rev. C **82** (2010) 25208
[264] G.T. Bodwin *et al.* arXiv:1307.7425 [hep-ph]
[265] COMPASS Collaboration (A. Austregesilo *et al.*), arXiv 1402.2170 [hep-ex]
[266] W. Boeglin *et al.*, Jefferson Lab experiment E12-10-003
[267] K. Slifer *et al.*, Jefferson Lab experiment E12-13-110
[268] J. Arrington *et al.*, Jefferson Lab experiment E12-10-008
[269] S. Gilad *et al.*, Jefferson Lab experiment E12-11-107
[270] J. Arrington *et al.*, Jefferson Lab experiment E12-11-112
[271] J. Arrington *et al.*, Jefferson Lab experiment E12-06-105
[272] P. Souder *et al.*, Jefferson Lab experiment E12-10-007
[273] I.C. Cloët *et al.*, Phy. Rev. Lett. **95** (2005) 052302
[274] W. Brooks *et al.*, Jefferson Lab experiment E12-14-001
[275] W. Brooks, *et al.*, Jefferson Lab Experiment E12-06-117.
[276] R.J. Holt and R. Gilman, Rept. Prog. Phys. **75** (2012) 086301
[277] S.J. Brodsky, C.R. Ji and  G.P. Lepage, Phys. Rev. Lett. **51** (1983) 83
[278] G.A. Miller, Phys. Rev. C **89** (2014) 045203
[279] S.J. Brodsky and A.H. Mueller, Phys. Lett. B **206** (1988) 685
[280] S Abrahamyan *et al.*, Phys. Rev. Lett. **108** (2012) 112502
[281] S. Riordan *et al.* Jefferson Lab experiment E12-12-004
[282] E.C. Aschenauer *et al.* arXiv:1501.01220 [nucl-ex]